\documentstyle[aps,twocolumn,eqsecnum]{revtex}
\begin{document}
\input psfig
\draft

\def\:={{\equiv}}

\wideabs{
\title{How unique is the expected stress-energy tensor of a massive scalar
field?}

\author{Wolfgang Tichy and \'Eanna \'E.\ Flanagan}
\address{Cornell University, Newman Laboratory, Ithaca, NY
14853-5001.}


\maketitle


\begin{abstract}
We show that the set of ambiguities in the renormalized expected
stress-energy tensor allowed by the Wald axioms is much larger for a
massive scalar field (an infinite number of free parameters) than for
a massless scalar field (two free parameters).
We also use the closed-time-path effective action formalism of Schwinger to
calculate the expected value of the stress-energy tensor in the
incoming vacuum state, for a massive scalar field,
on any spacetime which is a linear
perturbation off Minkowski spacetime.  This result generalizes an
earlier result of Horowitz and also Jordan in the massless case, and
can be used as a testbed for comparing different calculational methods.
\end{abstract}


}

\narrowtext

\section{Introduction and Summary}
\label{intro-sec}

\subsection{Background and Motivation}

In semiclassical gravity, a classical metric is coupled to quantum
fields according to the semiclassical Einstein equation
\begin{equation}
\label{Semiclass}
G_{ab}=8\pi G \, \langle \hat{T}_{ab} \rangle,
\end{equation}
where $G_{ab}$ is the Einstein tensor and $G$ is Newton's gravitational
constant.  This equation is usually postulated rather than derived
as there is no complete theory of quantum gravity from which it
could be derived, although several formal derivations have been given
\cite{anec1}.  There are several well-known difficulties associated
with the semiclassical theory.  First, there are difficulties
associated with the existence of unphysical, exponentially growing
``runaway'' solutions of Eq.\ (\ref{Semiclass}), which have not yet
been completely resolved \cite{anec}.  The second difficulty,
which is the subject of this paper, is the non-uniqueness of the
expected stress-energy tensor on the right hand side of
Eq.\ (\ref{Semiclass}).

For a scalar field, several methods have been suggested for calculating
the expected stress energy tensor.  These include
(i) the ``point splitting'' algorithm \cite{WaldQFT}, (ii) the
deWitt-Schwinger expansion method \cite{BirDav}, and (iii) the
closed-time-path or in-in effective action method
\cite{Schwing,J1,CampVar}.  There is no general agreement as to which
method is correct.  For example, it is claimed in Ref.\ \cite{WaldQFT}
that the deWitt-Schwinger method is invalid for a massive scalar field
since it does not have a regular limit as $m \to 0$, where $m$ is the
mass.

As is well known, a theorem of Wald \cite{Waldaxioms,WaldQFT} plays a
crucial role in this field.  The theorem states that if one has two
different prescriptions for obtaining stress tensors from metrics and
from quantum states, and if these prescriptions obey a certain set of
physically-motivated axioms, then the two prescriptions must agree up
to a local conserved tensor \cite{explainlocal}.  Thus, if
$\langle\hat{T}_{ab}\rangle$ and
$\langle\tilde{\hat{T}}_{ab}\rangle$ are
two different such prescriptions for computing the expected stress-energy
tensor, then the difference
\begin{equation}
\label{WaldThm}
t_{ab}\:= \langle\hat{T}_{ab}\rangle-\langle\tilde{\hat{T}}_{ab}\rangle
\end{equation}
must be a conserved local curvature tensor.

In the case of a massless scalar field, there is no natural mass scale in the
theory, and the following well-known argument based on
dimensional analysis shows that $\langle\hat{T}_{ab}\rangle$ is unique
up to a two parameter ambiguity.
Let us use units in which $\hbar = c=1$, but in which $G \ne 1$.
Then, there are only two
independent conserved local curvature
tensors with the appropriate dimensions of $(\mbox{mass})^4$, namely
\begin{eqnarray}
H^{(1)}_{ab}(x)=\frac{1}{\sqrt{-g}}\frac{\delta}{\delta g^{ab}(x)}
\int d^4 x' \sqrt{-g} R(x')^2
\label{H1gdef}
\end{eqnarray}
and
\begin{equation}
H^{(2)}_{ab}(x)=\frac{1}{\sqrt{-g}}\frac{\delta}{\delta g^{ab}(x)}
\int d^4 x' \sqrt{-g} R_{cd}(x')R^{cd}(x')   .
\label{H2gdef}
\end{equation}
Thus we must have
\begin{eqnarray}
\label{2-p-ambiguity}
t_{ab}=\alpha H^{(1)}_{ab} +\beta H^{(2)}_{ab} ,
\end{eqnarray}
where $\alpha$ and $\beta$ are two unknown dimensionless parameters.
Hence, the expected stress tensor $\langle\hat{T}_{ab}\rangle$ is
unique up to a two parameter ambiguity.

Consider now a massive scalar field.  In this case the above argument
fails, since there is a preferred mass scale present, namely the mass
$m$ of the field.  Using this mass scale one can construct local
conserved tensors of the form
\begin{eqnarray}
\label{try1}
\frac{1}{\sqrt{-g}}\frac{\delta}{\delta g^{ab}(x)}
\int d^4 x' \sqrt{-g} \, m^4 \Big[\frac{R(x')}{m^2}\Big]^n,
\end{eqnarray}
which have dimension $(\mbox{mass})^4$ and are thus possible
candidates for $t_{ab}$. Here $n$ can be any integer greater than 2.
Similar terms
can be constructed from the Ricci and Riemann tensors.  The
conventional view has been to exclude such terms as being unphysical,
since they diverge as $m \to 0$ (see, e.g., p.\ 90 of Ref.\
\cite{WaldQFT}).  Thus, the conventional view has been that the
allowed ambiguity for a massive field is no worse than that for a
massless field, namely the two parameter ambiguity
(\ref{2-p-ambiguity}).

\subsection{Ambiguity in expected stress tensor for a massive scalar
field}

The first main point of this paper is that the above conventional view
is unfounded.  This can be seen as follows.  Consider the local
conserved tensor
\begin{eqnarray}
\label{try2}
t_{ab}=\frac{1}{\sqrt{-g}}\frac{\delta}{\delta g^{ab}(x)}
\int d^4 x' \sqrt{-g} \, m^4 F\Big[\frac{R(x')}{m^2}\Big],
\end{eqnarray}
where $F(x)$ is any dimensionless function of a dimensionless argument
$x$.  In order for $t_{ab}$ to be acceptable on physical grounds as a
contribution to the expected stress tensor, it must satisfy the
requirements that
\begin{equation}
t_{ab} \to 0 \ \ \ \ {\rm as} \ \ \ \ m^2 \to 0
\label{physicalrequirement1}
\end{equation}
and
\begin{equation}
t_{ab} \to 0 \ \ \ \ {\rm as} \ \ \ \ R \to 0.
\label{physicalrequirement2}
\end{equation}
Now, the tensor (\ref{try2}) can be written as
\begin{eqnarray}
t_{ab} &=& - m^2 R_{ab} F^\prime(R/m^2) + m^4 F(R/m^2) g_{ab} /2
\nonumber \\ \mbox{} &&
+ F^{\prime \prime}(R/m^2) Y_{ab} + F^{(3)}(R/m^2) Z_{ab} / m^2,
\label{try2explicit}
\end{eqnarray}
where $Y_{ab}$ and $Z_{ab}$ are tensors constructed out of derivatives
of $R$, of dimension $({\rm mass})^{4}$ and $({\rm mass})^{6}$
respectively.  Suppose that we choose the function $F$ to be smooth,
to satisfy $F(0) = 0$, and to satisfy $F^{(j)}(x) x^{j - 2} \to 0$ as
$x \to \infty$ for $j = 0,1,2,3$, where $F^{(j)}(x)$ is the $j$th
derivative of $F$.   Examples of functions satisfying these
requirements are $F(x) = x^2 \exp(-x^2)$ and $F(x) = x^2 / (1 + x^4)$.
Then, the tensor (\ref{try2}) will satisfy the required properties
(\ref{physicalrequirement1}) and (\ref{physicalrequirement2}) \cite{note1}.

Note that when one expands the function $F(x)$ as a power series to obtain
\begin{eqnarray}
F\Big[\frac{R(x')}{m^2}\Big] =\sum_{n}a_{n}\Big[\frac{R(x')}{m^2}\Big]^n,
\end{eqnarray}
it can be seen that the tensor (\ref{try2}) contains terms of the form
(\ref{try1}), each
of which individually has unacceptable behavior as $m \to 0$.  However
the sum (\ref{try2}) of all these terms does have acceptable behavior
as $m \to 0$.  Note also that it is not possible to exclude terms of
the form (\ref{try2}) by postulating, as an additional axiom, that the
stress tensor be an analytic function of $m^2$, since for example the
choice $F(x) = x^2 / (1 + x^4)$ yields a local conserved tensor
$t_{ab}$ which is an analytic function of $m^2$ in an open
neighborhood of the real axis in the complex $m^2$ plane \cite{note4}.

The ambiguity $t_{ab}$ in the stress-energy tensor allowed by the Wald
axioms is therefore much worse in the massive case than in the
massless case.  It is an infinite parameter ambiguity --- one can
specify a free function $F(x)$ --- rather than a two parameter
ambiguity \cite{notepointsplit}.
Of course, it is still possible that the various
conventional calculational methods still agree to within the two
parameter ambiguity (\ref{2-p-ambiguity}) \cite{Moretti}.  However,
there is no
guarantee that this should be the case.  Therefore it would be
worthwhile to find some additional axiom or physical principle, to
augment the Wald axioms,  that would further pin down the stress
tensor in the massive case \cite{note22}.

\subsection{Nearly flat spacetimes}

Spacetimes which are linear perturbations off Minkowski spacetime
form a useful testbed in which to probe these issues \cite{note2}.
The second principle purpose of this paper is to explicitly calculate
the renormalized stress tensor of a massive scalar field in such
spacetimes, using the closed-time-path or in-in effective action formalism
\cite{Schwing,J1,CampVar}.  If our calculation is repeated using the
point-splitting or deWitt-Schwinger methods, then it will be possible
to compare the predictions of the different methods.

Now the Wald axioms imply that any prescription for calculating the
stress tensor is determined by specifying the expected value of the
stress tensor in the incoming vacuum state $\left| 0,{\rm in}
\right>$.   The expected value in any other state is then uniquely
determined \cite{anec}.  Therefore, it suffices to consider the
expected value of the stress tensor in the incoming vacuum state.
In the massless case, calculations of the in-in expected stress tensor
have already been performed using several different methods
[see Horowitz \cite{Horowitz} and Jordan \cite{J2}]. The results of
of these different calculations agree up to the two parameter
ambiguity (\ref{2-p-ambiguity}), as they must according to Wald's
theorem.

The result we obtain from the in-in effective action formalism
\cite{Schwing,J1,CampVar} is [cf.\ Eq.\ (\ref{Tab-in-in}) below]
\begin{eqnarray}
\label{intro-Tab-in-in}
\left< 0,{\rm in} \right| {\hat T}_{ab}(x) && \left| 0,{\rm in}
\right>  =
\alpha \dot{H}^{(1)}_{ab}(x)+ \beta \dot{H}^{(2)}_{ab}(x) \nonumber \\
&&-\frac{1}{256\pi^2}\int d^4 x'
      \Big[\dot{H}^{(1)}_{ab}(x')T_1 (x-x') \nonumber \\
&& \ \             +\dot{H}^{(2)}_{ab}(x')T_2 (x-x')\Big].
\end{eqnarray}
Here it is assumed that the metric tensor is of the form
\begin{equation}
g_{ab} = \eta_{ab} + h_{ab},
\label{nearly-flat}
\end{equation}
where $\eta_{ab}$ is a flat, Minkowski metric, and that the
coordinates $x^a$ and $x^{\prime\, a}$ are Lorentzian coordinates with
respect to $\eta_{ab}$.  Also $\alpha$ and $\beta$ are
arbitrary dimensionless constants,
the distributions $T_1 (x-x')$ and $T_2 (x-x')$ are defined by
Eqs.\ (\ref{tildeT1}) and (\ref{tildeT2}) below, and ${\dot
H}^{(1)}_{ab}$ and  ${\dot H}^{(2)}_{ab}$ are linearized versions of
the local conserved curvature tensors (\ref{H1gdef}) and
(\ref{H2gdef}).

The stress-energy tensor (\ref{intro-Tab-in-in})
is causal, as it must be, and reduces to the known result of the
massless case \cite{Horowitz,J2} in the limit $m \to 0$.
Furthermore it is not a smooth function of $m^2$ at $m^2 =0$.
The calculational method we use also automatically yields two
undetermined parameters $\alpha$ and $\beta$, so the result
(\ref{intro-Tab-in-in}) explicitly exhibits the two parameter
ambiguity, just as in the massless case \cite{Horowitz}.

\subsection{Organization of this paper}

Section \ref{QM-PI} reviews the in-out and in-in effective action
formalisms.  In Sec.\ \ref{Calc} we calculate
the in-in and in-out effective actions of a massive scalar field
propagating on a spacetime which is a linear perturbation off
Minkowski spacetime.
In  Sec. \ref{stress-energy}  we find the  expected stress-energy in the
incoming vacuum state, and  then  in Sec.\  \ref{Properties} we discuss
its properties.  Sec. \ref{Conclusion} summarizes our results.

Throughout we use units in which $\hbar = c =1$, and use the
metric signature and sign conventions of Misner, Thorne and
Wheeler \cite{MTW}.  Further notational conventions are given in
Appendix \ref{Notation}.

\section{The in-out and in-in path integral formalisms}
\label{QM-PI}

In this section we review both the standard in-out
path integral formalism of quantum field theory (see, e.g., Refs.\
\cite{Peskin,BirDav}), as applied to curved spacetimes, and the
modified in-in formalism due to Schwinger \cite{Schwing}.
This in-in or closed time path method was later adapted to curved
spacetimes by Jordan \cite{J1,CampVar}, and has been extensively
explored by Hu \cite{Hu}. 
Our presentation of the in-in method in Sec.\ \ref{QM-PI-in-in} below
differs from that of Refs.\ \cite{J1,CampVar} in that all the
fundamental definitions are explicitly coordinate independent.

\subsection{The classical theory}
\label{classic}

We consider a massive, minimally coupled scalar field $\phi$ for which
the Einstein-Klein-Gordon action is
\begin{eqnarray}
\label{class-action}
S[g^{ab},\phi]=S_{g}[g^{ab}]+S_{m}[g^{ab},\phi],
\end{eqnarray}
where
\begin{eqnarray}
\label{class-grav}
S_{g}[g^{ab}]= 2\mu_p^2 \int \sqrt{-g} R d^n x
\end{eqnarray}
and
\begin{eqnarray}
\label{class-matter}
S_{m}[g^{ab},&&\phi]  \nonumber \\
&&=-\frac{1}{2}\int\sqrt{-g} (g^{ab}\nabla_a
\phi\nabla_b \phi +m^2\phi^2)d^n x.
\end{eqnarray}
Here $m$ is the mass of the scalar field, $\mu_p^2=(32\pi G)^{-1}$ is
the square of the Planck mass, and
$n$ is the number of spacetime dimensions (we shall be using
the dimensional regularization scheme below).
The corresponding equations of motion are
$4\mu_p^2 G_{ab}=T_{ab}
$
and
$(\Box - m^{2} )\phi=0$,
where the stress-energy tensor is
\begin{eqnarray}
\label{T-class}
T_{ab}=
&=&\frac{-2}{\sqrt{-g}} \frac{\delta}{\delta
         g^{ab}}S_{m}[g^{ab},\phi] \nonumber \\
         &=&
\nabla_a \phi \nabla_b \phi-\frac{1}{2}g_{ab}\nabla_c \phi \nabla^c \phi
         -\frac{1}{2}g_{ab}m^{2}\phi^{2} .
\end{eqnarray}

\subsection{In-out formalism}

We assume that the metric $g_{ab}$ is asymptotically static at early and late
times so the incoming and outgoing vacuum states $
|0,\mbox{in}\rangle$ and $|0,\mbox{out}\rangle$
are well defined.  
In the usual way we define the generating functional
\begin{equation}
e^{iW[g^{ab},J]}\, \:= \, \langle0,\mbox{out}|0,\mbox{in}\rangle_{J},
\end{equation}
where the subscript $J$ on the right hand side indicates that a source
term
\begin{equation}
\langle J, \phi \rangle \equiv \int d^n x \, \sqrt{-g(x)} \, J(x)\phi(x)
\label{product-notation}
\end{equation}
has been added to the action.  It can
be shown that
$e^{iW[g^{ab},J]}$ has the path integral representation
\begin{equation}
\label{in-out-W}
e^{iW[g^{ab},J]}=\int D\phi \ e^{i(S_{m}[g^{ab},\phi]+\langle J,
\phi\rangle)} ,
\end{equation}
and that time-ordered matrix elements are given by
\begin{eqnarray}
\label{in-out-elem}
\langle0,\mbox{out}|T\hat{\phi}(x)\hat{\phi}(y)|0,\mbox{in}\rangle_J
&=&\int D\phi \ \phi(x)\phi(y) \nonumber \\
&& \times e^{i(S^{m}[g^{ab},\phi]+\langle J,\phi\rangle )}
. \nonumber \\
\end{eqnarray}
In Eqs.\ (\ref{in-out-W}) and (\ref{in-out-elem}), the usual
boundary conditions on the path integral are assumed, namely that
$\phi$ is purely negative frequency ($\propto e^{i \omega t}$ with
$\omega >0$) at early times and positive
frequency at late times, or equivalently, that the mass squared
parameter is understood to have a small negative imaginary part, $m^2
\to m^2 - i \epsilon$.  
{}From Eqs.\ (\ref{in-out-elem}) and (\ref{T-class}) it
follows that
\begin{eqnarray}
\frac{-2}{\sqrt{-g}}\frac{\delta W[g^{ab},0]}{\delta g^{ab}} &=&
-i e^{-iW[g^{ab},0]} \int D\phi \ \nonumber \\
&& \times \, i\frac{-2}{\sqrt{-g}}\frac{\delta}{\delta
g^{ab}}S_{m}[g^{ab},\phi]
e^{iS_{m}[g^{ab},\phi]} \nonumber \\
&=&\frac{\langle0,\mbox{out}|\hat{T}_{ab}|0,\mbox{in}\rangle}
{\langle0,\mbox{out}|0,\mbox{in}\rangle}.
\label{nn1}
\end{eqnarray}

The
effective action is defined in the usual way as a Legendre transform
of the generating functional:
\begin{eqnarray}
\label{in-out-effaction}
\Gamma_{m}[g^{ab},\bar{\phi}] &\:= & W[g^{ab},J]-\langle J,\bar{\phi}\rangle ,
\end{eqnarray}
where
\begin{eqnarray}
\label{def-phi-bar}
\bar{\phi} &=& {1 \over \sqrt{-g}} \frac{\delta W[g^{ab},J]}{\delta J}
=\langle0,\mbox{out}|\hat{\phi}|0,\mbox{in}\rangle_{J} .
\end{eqnarray}
Here and henceforth the subscript $m$ in $\Gamma_m$ indicates that the
classical action from which $\Gamma_m$ is computed [Eq.\
(\ref{in-out-W}) above] includes only the matter part $S_m$ and not
the gravitational part $S_g$.  
Combining Eqs.\ (\ref{nn1}) -- (\ref{def-phi-bar}) we now obtain
\begin{eqnarray}
\label{Tab-in-out-def}
\left.\frac{-2}{\sqrt{-g}} \frac{\delta \Gamma_{m}[g^{ab},\bar{\phi}]}{\delta
g^{ab}} \right|_{{\bar \phi} = {\bar \phi}[g^{ab}]}
&=&\frac{\langle0,\mbox{out}|\hat{T}_{ab}|0,\mbox{in}\rangle}
        {\langle0,\mbox{out}|0,\mbox{in}\rangle}.
\end{eqnarray}
Here the right hand side is a functional only of the metric $g^{ab}$,
while on the left hand side ${\bar \phi}[g^{ab}]$ is
the solution to Eq.\ (\ref{def-phi-bar}) at $J=0$.

Since the action (\ref{class-matter}) is quadratic in $\phi$, it is
straightforward to compute the effective action exactly.  The result
is the standard, formal, expression \cite{Peskin0}
\begin{equation}
\Gamma_m[g^{ab},{\bar \phi}] = S_m[g^{ab},{\bar \phi}] + {i \over 2}
\, {\rm tr} \,\,  {\rm ln} \, {\bar A},
\label{Gammam-in-out-formal}
\end{equation}
where ${\bar A}$ is the operator given by ${\bar A} \phi = (\Box - m^2 + i \epsilon) \phi$.
The operator ${\bar A}$ is the natural operator associated with
the quadratic form $S_m$ and with the covariant inner product on functions
on spacetime
\begin{equation}
\label{innerproduct-in-out-covariant}
\left< f,h \right> \, \equiv \, \int d^n x \, \sqrt{-g(x)} \, f(x)^*
h(x).
\end{equation}
The reason the inner product (\ref{innerproduct-in-out-covariant}) is
the appropriate inner product is that the measure $D \phi$ in Eq.\
(\ref{in-out-W}) is determined by the metric $g^{ab}$ \cite{measure}.


Finally, we define the quantity $\Gamma[g^{ab},{\bar \phi}]$
(as opposed to $\Gamma_m[g^{ab},{\bar \phi}]$) to be the effective
action obtained when one starts from the full action
(\ref{class-action}) rather than just the matter part
(\ref{class-matter}).  It is clear that
\begin{eqnarray}
\Gamma[g^{ab},\bar{\phi}]=\Gamma_{m}[g^{ab},\bar{\phi}]
+S_{g}[g^{ab}].
\label{Gamma-in-out-def}
\end{eqnarray}

\subsection{In-in formalism}
\label{QM-PI-in-in}

We introduce the generating functional
\FL
\begin{equation}
\label{W-in-in-def}
e^{iW[g^{ab}_{+},g^{ab}_{-},J_+,J_-]}\, \:= \int D\alpha \,
\langle0,\mbox{in}|\alpha,T\rangle_{J_-}
\langle\alpha,T|0,\mbox{in}\rangle_{J_+} ,
\end{equation}
which depends on two independent sources $J_+(x)$ and $J_-(x)$, as
well as two metrics
$g^{ab}_+$ and $g^{ab}_-$.  Equation (\ref{W-in-in-def}) includes an
integration over a complete set of
states $|\alpha,T\rangle$ on the hypersurface $x^0 = T$ at some future
time $T$.  We assume that $g_+^{ab} = g_-^{ab}$ for $x^0 \ge T$.
Each of the
matrix elements in Eq.\ (\ref{W-in-in-def}) can be expressed as path
integrals in the usual way:
\begin{equation}
\label{stPI}
\langle\alpha,T|0,\mbox{in}\rangle_{J_{\pm}} =\int D\phi_{\pm}
e^{i(S_{m}[g^{ab}_{\pm},\phi_{\pm}]+\langle J_{\pm},\phi_{\pm} \rangle_\pm)},
\end{equation}
where the measure $D \phi_+$ and inner product $\langle \ldots \rangle_+$
are determined by the metric $g_+^{ab}$,
the measure $D \phi_-$ and the inner product $\langle \ldots \rangle_-$
by the metric $g_-^{ab}$.
Combining Eqs.\ (\ref{W-in-in-def}) and (\ref{stPI}) yields
\begin{eqnarray}
e^{iW[g^{ab}_{+},g^{ab}_{-},J_+,J_-]}&&=\int D\alpha \nonumber \\
&& \times \int D\phi_{-}
e^{-i(S_{m}^{*}[g^{ab}_{-},\phi_{-}]+\langle J_{-}, \phi_{-}\rangle_- )}
\nonumber \\
&& \times  \int D\phi_{+} e^{i(S_{m}[g^{ab}_{+},\phi_{+}]+ \langle
J_{+}, \phi_{+} \rangle_+)} ,
\label{www}
\end{eqnarray}
with the boundary condition that $\phi_{+}=\phi_{-}=\alpha$ on the
hypersurface given by $x^0=T$.  Another boundary condition, needed to
assure convergence of the path integrals, is that
$\phi_{+}$ be purely negative frequency and $\phi_-$ be purely
negative frequency at early times, or equivalently that $m^2$ be
interpreted as $m^2 - i \epsilon$ in the action (\ref{class-matter}).
{}From now on we assume the second of these.
We can rewrite the generating functional (\ref{www}) as
\begin{eqnarray}
\label{in-in-W}
&e&^{iW[g^{ab}_{+},g^{ab}_{-},J_{+},J_{-}]}  \nonumber \\
&&=\int D\phi_{+}D\phi_{-}
{\rm exp}\, i\, \Big\{ S_{m}[g^{ab}_{+},\phi_{+}]+ \langle
J_{+},\phi_{+} \rangle_+ \nonumber \\ &&
-S_{m}^{*}[g^{ab}_{-},\phi_{-}]- \langle J_{-},\phi_{-} \rangle_- \Big\}
 , \nonumber \\
\end{eqnarray}
where the integral $\int D \alpha$ is now included in the integration
over $\phi_+$ and $\phi_-$, and the boundary condition is that
$\phi_{+}=\phi_{-}$ on the hypersurface given by $x^0=T$.
Below we will be taking the limit $T \rightarrow \infty$.
{}From Eq.\ (\ref{in-in-W}) it follows that 
\begin{eqnarray}
\label{Tab-from-W}
\frac{-2}{\sqrt{-g_{+}}}
\frac{\delta W[g^{ab}_{+},g^{ab}_{-},0,0]}{\delta g^{ab}_{+}}
&&\Big|_{g^{ab}_{+}=g^{ab}_{-}=g^{ab}} \nonumber \\ &&
=\langle0,\mbox{in}|\hat{T}_{ab}|0,\mbox{in}\rangle.
\end{eqnarray}

The effective action is defined to be a Legendre
transform of the generating functional as before:
\begin{eqnarray}
\label{in-in-effaction}
\Gamma_{m}&[&g^{ab}_{+},g^{ab}_{-},\bar{\phi}_{+},\bar{\phi}_{-}]
\\
&&\:= \, W[g^{ab}_{+},g^{ab}_{-},J_{+},J_{-}]
-\langle J_{+},\bar{\phi}_{+}\rangle_+ + \langle J_{-}, \bar{\phi}_{-}
\rangle_- ,  \nonumber
\end{eqnarray}
where
\begin{eqnarray}
\label{phi-bar-def}
\bar{\phi}_{\pm}[g_+^{ab},g_-^{ab},J_+,J_-]
&=& \pm{1 \over \sqrt{g_\pm}} \, \frac{\delta
W[g^{ab}_{+},g^{ab}_{-},J_{+},J_{-}]}{\delta J_{\pm}}, \nonumber \\
\end{eqnarray}
and where we use the shorthand notation (\ref{product-notation}).
Note that when $g_+^{ab} = g_-^{ab}$ and $J_+ = J_-$, we find from
Eqs.\ (\ref{in-in-W}) and (\ref{in-in-effaction}) that
\begin{eqnarray}
\bar{\phi}_{\pm}[g^{ab},g^{ab},J,J]
&=& \langle0,\mbox{in}|\hat{\phi}|0,\mbox{in}\rangle_J.
\label{vev}
\end{eqnarray}
{}From Eqs.\ (\ref{Tab-from-W}) and (\ref{in-in-effaction})
we obtain the expected stress-energy tensor in the in-coming
vacuum state:
\begin{eqnarray}
\label{Tab-from-Gamma}
T_{ab\,{\rm in-in}}&\:= &
\langle0,\mbox{in}|\hat{T}_{ab}|0,\mbox{in}\rangle
\nonumber \\ &=&
\frac{-2}{\sqrt{-g_+}} \frac{\delta
\Gamma_{m}} {\delta
g^{ab}_{+}} \Bigg|_{\bar{\phi}_{\pm}=\bar{\phi}[g^{ab}] \ , \
g^{ab}_{\pm}=g^{ab}
}.
\end{eqnarray}
Here ${\bar \phi}[g^{ab}]$ is given by Eq.\ (\ref{phi-bar-def}) at
$J_+ = J_- = 0$ and $g_+^{ab} = g_-^{ab} = g^{ab}$:
\begin{equation}
{\bar \phi}[g^{ab}] = {\bar \phi}_+[g^{ab},g^{ab},0,0].
\label{bar-phi-value}
\end{equation}
{}From Eq.\ (\ref{vev}) it can be seen that ${\bar \phi}[g^{ab}]$ is the
expected value of the field in the incoming vacuum state.
Equation (\ref{Tab-from-Gamma}) is the formula we will use below to
compute the stress tensor.

We introduce the shorthand notations $\phi_s = (\phi_+, \phi_-)$, 
\begin{equation}
\label{hatJ-def}
{\hat J}_{s} = \Bigg(
\begin{array}{l}
+J_{+} \\ -J_{-} \\
\end{array}
\Bigg),
\end{equation}
and
\begin{equation}
S_{m}[g_s^{ab},\phi_{s}] =S_{m}[g_+^{ab},\phi_{+}]-S_{m}^*[g_-^{ab},\phi_{-}],
\label{Sm-in-in-def}
\end{equation}
where the index $s$ takes the values $+$ and $-$.  The generating
functional (\ref{in-in-W}) can be rewritten using these notations as
\begin{equation}
\label{exp-W}
e^{iW[g_s^{ab},{\hat J}_s]}=\int D\phi_{r}\
e^{i(S_{m}[g_s^{ab},\phi_{s}]+\langle {\hat J}_t, \phi_{t} \rangle_t) },
\end{equation}
where a sum over the repeated index $t$ is understood.

Next, we derive the analog of the formal expression
(\ref{Gammam-in-out-formal}) for the in-in effective action.  We
define the operator ${\bar A}$ on pairs of functions $\phi_s =
(\phi_+,\phi_-)$ by
\begin{equation}
({\bar A}\,\phi)_s(x) = \int d^n y \, \sqrt{-g_t(y)} {\bar A}_{st}(x,y)
\, \phi_t(y),
\label{A-rep-covariant}
\end{equation}
where
\begin{eqnarray}
{\bar A}_{st}(x,y)\:=  {1 \over \sqrt{g_s(x) g_t(y)}} \frac{\delta^2
S_{m}[g_r^{ab},\phi_r]} {\delta\phi_{s}(x) \delta\phi_{t}(y) },
\label{A-in-in-def-covariant}
\end{eqnarray}
which can be written using Eqs.\ (\ref{class-matter}) and
(\ref{Sm-in-in-def}) as
\begin{eqnarray}
\label{kernel1}
{\bar A}_{st}(x,y)&=&
\label{A-in-in-formula-covariant}
\delta^{n} (x-y)
\Bigg[
\begin{array}{cc}
{ \Box_{+,x} -m^2+i\epsilon \over \sqrt{-g_+(x)}}
& 0 \\ 0 & {-\Box_{-,x} +m^2+i\epsilon \over \sqrt{-g_-(x)}}\\
\end{array}
\Bigg]. \nonumber \\
\end{eqnarray}
Here $\Box_{+,x}$ denotes the wave operator associated with the metric
$g_+^{ab}$ acting on the coordinates $x = x^a$, and similarly for
$\Box_{-,x}$.  Using Eqs.\ (\ref{in-in-effaction}), (\ref{exp-W}) and
(\ref{A-in-in-def-covariant}), we then find
the following analog of Eq.\ (\ref{Gammam-in-out-formal})
\begin{eqnarray}
\label{Gamma-approx}
\Gamma_{m}[g^{ab}_{r},\bar{\phi}_{r}] &=&S_{m}[g^{ab}_{r},\bar{\phi}_{r}]
+\frac{i}{2} {\rm Tr} \,\, {\rm ln} \left({\bar A}_{st} \right),
\end{eqnarray}
where ${\rm Tr}$ denotes the appropriate trace over both spacetime
variables and over the indices $s,t$.
The operator ${\bar A}$ is the natural operator associated with
the quadratic form (\ref{Sm-in-in-def}) and with the covariant inner product
on pairs of functions $(f_+,f_-)$ given by
\begin{equation}
\label{innerproduct-in-in-covariant}
\left< (f_+,f_-), (h_+,h_-) \right>  \equiv  \sum_{s = +,-}
\int d^n x  \sqrt{-g_s(x)}  f_s(x)^* h_s(x).
\end{equation}
The reason the inner product (\ref{innerproduct-in-in-covariant}) is
the appropriate inner product is that the measures $D \phi_+$ and $D
\phi_-$ in Eq.\ (\ref{in-in-W}) are determined by the metrics $g_+^{ab}$
and $g_-^{ab}$ respectively.

In our perturbative computations below, we shall derive an expression
for the effective action in terms of a series of products of
operators.  For that purpose, it will be convenient to use the 
Hilbert space structure associated with the coordinate
dependent inner product 
\begin{equation}
\label{innerproduct-in-in}
\left< (f_+,f_-), (h_+,h_-) \right>^{\rm c} \, \equiv \, \sum_{s = +,-} \,
\int d^n x \, f_s(x)^* h_s(x),
\end{equation}
instead of that associated with the covariant inner product
(\ref{innerproduct-in-in-covariant}).  We will always choose the
coordinate system $x^a$ 
appearing in Eq.\ (\ref{innerproduct-in-in}) to be a Lorentzian
coordinate system associated with the flat metric $\eta_{ab}$.

Finally, we define the quantity $\Gamma[g_+^{ab},g_-^{ab},{\bar
\phi}_+,{\bar \phi}_-]$ (as opposed to $\Gamma_m$) to be the effective
action obtained when one starts from the full action
(\ref{class-action}) rather than just the matter part
(\ref{class-matter}).  It is clear that
\begin{eqnarray}
\Gamma[g_+^{ab},g_-^{ab},\bar{\phi}_+,{\bar \phi}_-]&=&
\Gamma_m[g_+^{ab},g_-^{ab},\bar{\phi}_+,{\bar \phi}_-] \nonumber \\
&&
+S_{g}[g_+^{ab}] -S_{g}[g_-^{ab}].
\label{Gamma-in-in-def}
\end{eqnarray}
The semiclassical equations of motion are given by
\begin{equation}
{\delta \Gamma \over \delta g_+^{ab}} \Bigg|_{\bar{\phi}_\pm =
\bar{\phi} \ , \ g^{ab}_{\pm}=g^{ab}}
=
{\delta \Gamma \over \delta {\bar \phi}_+} \Bigg|_{
\bar{\phi}_\pm = \bar{\phi} \ , \ g^{ab}_{\pm}=g^{ab}
}
=0.
\label{Gamma-eqn-of-motion}
\end{equation}
The second of these equations is automatically solved when we
choose ${\bar \phi} = {\bar \phi}[g^{ab}]$, cf.\ Eq.\
(\ref{bar-phi-value}) above, 
corresponding to the incoming vacuum state.  Thus the equation of
motion reduces to
\begin{eqnarray}
\label{Gamma-eqn-of-motion-1}
{\delta \Gamma \over \delta g_+^{ab}} \Bigg|_{
\bar{\phi}_\pm = \bar{\phi}[g^{ab}] \ , \ g^{ab}_{\pm}=g^{ab} 
 }
=0.
\end{eqnarray}

\section{THE EFFECTIVE ACTION FOR NEARLY FLAT SPACETIMES}
\label{Calc}

In this section we specialize to almost flat spacetimes of the form
(\ref{nearly-flat}), and calculate the in-out effective action
(\ref{Gamma-in-out-def}) and
the in-in effective action (\ref{Gamma-in-in-def}) as series expansions
in powers of the metric perturbation $h_{ab}$.  We use the methods of
Hartle and Horowitz \cite{HarHor} and of Jordan
\cite{J1,J2}, who performed similar calculations in the massless case.

We start by further simplifying Eq.\ (\ref{Gamma-approx}).  Note that
\begin{eqnarray}
\label{314}
\frac{\delta}{\delta g_{+}^{ab}} \mbox{Tr} \,\, \mbox{ln} \left[ {\bar
A}_{st} \right]
&=&\mbox{Tr}\left({\bar A}_{sr}^{-1}
\frac{\delta {\bar A}_{rt}}{\delta g_{+}^{ab}}\right) \\
&=&\mbox{tr}\left({\bar A}_{++}^{-1}
\frac{\delta {\bar A}_{++}}{\delta g_{+}^{ab}}\right) ,
\label{315}
\end{eqnarray}
since in the sum over $r$ in Eq.\ (\ref{314}),
only ${\bar A}_{++}$ depends on $g^{ab}_{+}$.
Here $\mbox{tr}$ denotes a trace over the spacetime variables $x,y$
only, not including a sum over the $s,t$ indices.  Combining Eqs.\
(\ref{Gamma-approx}) and (\ref{315})
we now obtain
\begin{eqnarray}
\label{Gamma-approx2}
\Gamma_{m}[g^{ab}_{r},\bar{\phi}_{r}] &=&
S_{m}[g^{ab}_{r},\bar{\phi}_{r}] +\frac{i}{2} \mbox{tr} \,\, \mbox{ln}
\, {\bar A}_{++} + F_1[g_-^{ab}],
\end{eqnarray}
where $F_1[g_-^{ab}]$ is some functional of $g_-^{ab}$.  This term
$F_1[g_-^{ab}]$ will not contribute to the
functional derivative in Eq.\ (\ref{Tab-from-Gamma}) and hence will
not contribute to the in-in expected stress tensor.

Next, we define the coordinate dependent operator $A = A_{rs}$ acting
on pairs of functions $(\phi_+,\phi_-)$ by
\begin{equation}
\left( A \phi \right)_r(x) = \sqrt{- g_r(x) } \left( {\bar A} \phi
\right)_r(x).
\label{A-in-in-def-0}
\end{equation}
Then from Eqs.\ (\ref{A-rep-covariant}) and
(\ref{A-in-in-formula-covariant})
the kernel $A_{rs}(x,y)$ of $A$ with respect to the coordinate
dependent inner product (\ref{innerproduct-in-in}) is given by
\begin{equation}
A_{rs}(x,y) = \sqrt{-g_r(x)} \sqrt{-g_s(y)} {\bar A}_{rs}(x,y).
\label{A-in-in-def-1}
\end{equation}
[Here ${\bar A}_{rs}(x,y)$ is the kernel of the operator ${\bar A}$
with respect to the inner product
(\ref{innerproduct-in-in-covariant}), given by Eq.\
(\ref{A-in-in-formula-covariant})]. 
Combining Eqs.\ (\ref{A-in-in-formula-covariant}) and
(\ref{Gamma-approx2}) -- (\ref{A-in-in-def-1}) 
now yields
\begin{eqnarray}
\label{Gamma-approx2a}
\Gamma_{m}[g^{ab}_{r},\bar{\phi}_{r}] &=&
S_{m}[g^{ab}_{r},\bar{\phi}_{r}] + F_1[g_-^{ab}] \nonumber \\ &&
+\frac{i}{2} \mbox{tr} \,\, \mbox{ln}
\, A_{++} + F_2[{\rm det}(g_+)],
\end{eqnarray}
where $F_2[{\rm det}(g_+)]$ is some functional of the determinant
${\rm det} [g_+^{ab}(x)]$.
When we take the variational derivative of $\Gamma_m$ in Eq.\
(\ref{Tab-from-Gamma}) to calculate the stress tensor, the term $F_2$
will contribute a term proportional to the metric $g_{ab}$ and hence will
contribute only to the renormalization of the cosmological constant.

We define the propagator ${\bar G}_{st}$ to be the inverse of the
operator ${\bar A}_{st}$.  Its kernel ${\bar G}_{st}(x,y)$ with
respect to the inner product (\ref{innerproduct-in-in-covariant}) is
given by
\begin{equation}
\label{prop-def-cov}
\int d^n x' \sqrt{-g_s(x')} \, {\bar A}_{rs}(x,x') {\bar G}_{st}(x',y) =
-{ \delta^n (x-y) \over \sqrt{-g_t(y)} } \, \delta_{rt}.
\end{equation}
Note that the operation of taking the inverse is unique by virtue of
the boundary conditions imposed on the path integrals and the fact
that the mass squared parameter is assumed to have a small negative
imaginary part (see Appendix \ref{App-invers}).
We also define a coordinate dependent operator $G_{st}$ to be the
inverse of the operator $A_{st}$; its kernel $G_{st}(x,y)$ with
respect to the inner product (\ref{innerproduct-in-in}) is
given by
\begin{equation}
\label{prop-def}
\int d^n x'  \, A_{rs}(x,x') G_{st}(x',y) = - \delta^n
(x-y) \, \delta_{rt}.
\end{equation}
{}From Eqs.\ (\ref{A-in-in-def-1}),
(\ref{prop-def-cov}) and (\ref{prop-def}) it follows that
$G_{rs}(x,y) = {\bar G}_{rs}(x,y)$, that is, the above two kernels
coincide.

\subsection{Perturbation expansion for the in-in effective action}

We now expand the operator $A_{st}$ as
\begin{eqnarray}
A_{rs}(x,x') = A_{rs}^{0}(x,x') + V_{rs}(x,x'),
\label{A-expansion}
\end{eqnarray}
where
\begin{eqnarray}
V_{rs}(x,x') = V_{rs}^{(1)}(x,x') + V_{rs}^{(2)}(x,x') + \ldots
\end{eqnarray}
Here $A^0_{rs}$ is the Minkowski spacetime operator, and the
terms $V^{(1)}$ and $V^{(2)}$ are the pieces of $A_{rs}$ that are
linear and quadratic in the metric perturbation $h_{ab}$, respectively [see
Eqs.\ (\ref{V1}) and (\ref{V2}) below].
Note that from Eqs.\ (\ref{A-in-in-formula-covariant}),
(\ref{A-in-in-def-1}) and (\ref{A-expansion}) it 
follows that the operator $V_{rs}$ is diagonal in the indices $r$ and
$s$ and is of the form
\begin{eqnarray}
\label{V-defn}
V_{rs}[g_+^{ab},g_-^{ab}]
&=& \Bigg[
\begin{array}{cc}
V_{++}[g_+^{ab},g_-^{ab}]  & V_{+-}[g_+^{ab},g_-^{ab}]  \\
V_{-+}[g_+^{ab},g_-^{ab}]  & V_{--}[g_+^{ab},g_-^{ab}] \\
\end{array}
\Bigg]  \nonumber \\
&=& \Bigg[
\begin{array}{cc}
V[g_+^{ab}] & 0 \\ 0 & -V[g_-^{ab}]^* \\
\end{array}
\Bigg],
\end{eqnarray}
for some functional $V = V[g^{ab}]$.

We similarly expand the propagator $G_{st}$
as
\begin{equation}
G_{st} = G_{st}^0 + G_{st}^{(1)} + G_{st}^{(2)} + \ldots
\label{G-expansion}
\end{equation}
The Minkowski spacetime propagator $G_{st}^0$ can be obtained by
combining Eqs.\ 
(\ref{A-in-in-formula-covariant}), (\ref{A-in-in-def-1})
and (\ref{prop-def}) and using
$g_{ab} = \eta_{ab}$.  In Appendix \ref{App-invers} we show that this
yields
\begin{eqnarray}
\label{free-prop-ma}
G_{st}^{0}(x,y)&=& \Bigg[
\begin{array}{cc}
G_{++}^{0}(x,y) & G_{+-}^{0}(x,y) \\ G_{-+}^{0}(x,y) & G_{--}^{0}(x,y)\\
\end{array}
\Bigg]  \nonumber \\
&=& \Bigg[
\begin{array}{cc}
 G(x-y) & \Delta^{+}(x-y) \\ -\Delta^{+}(x-y)^* & -G(x-y)^*\\
\end{array}
\Bigg]  ,
\end{eqnarray}
where
\begin{equation}
G(x)= \int \frac{d^n
p}{(2\pi)^n}\frac{e^{ipx}}{p^2+m^2-i\epsilon}
\label{Feynman-prop}
\end{equation}
is the free Feynman propagator and
\begin{equation}
\label{wighty}
\Delta^{+}(x)= 2\pi i\int \frac{d^n p}{(2\pi)^n} e^{ipx} \delta
(p^2+m^2)\Theta{(-p^0)}
\end{equation}
is the positive Wightman function.

Next, by combining the expansions (\ref{A-expansion}) and
(\ref{G-expansion}) together with
the definition (\ref{prop-def}), we find the following expression for
the logarithmic term appearing in the effective action
(\ref{Gamma-approx2}) (see Appendix \ref{expandprop}):
\begin{eqnarray}
\label{log-G++}
\mbox{ln}[G_{++}]
&=&\mbox{ln}[G_{++}^{0}]
+V_{++}^{(1)}G_{++}^{0} +V_{++}^{(2)}G_{++}^{0} \nonumber \\
&&+\frac{1}{2}V_{++}^{(1)}G_{++}^{0}
V_{++}^{(1)}G_{++}^{0} +V_{++}^{(1)}G_{+-}^{0}
V_{--}^{(1)} G_{-+}^{0} \nonumber \\
&&+O(h^3).
\end{eqnarray}
The products on the right hand side of Eq.\ (\ref{log-G++}) are
operator products, where the kernels are understood to refer to the
inner product (\ref{innerproduct-in-in}), so that, for example,
$[V_{++}^{(1)}G_{++}^{0}](x,y)  \equiv \int d^n y'
V_{++}^{(1)}(x,y')G_{++}^{0}(y',y)$.
Combining Eqs.\ (\ref{log-G++}) and (\ref{Gamma-approx2})
we finally obtain the perturbation expansion of the in-in effective
action
\FL
\begin{eqnarray}
\label{Gamma-approx-fin}
\Gamma_{m}[g^{ab}_{r},\bar{\phi}_{r}]
&=&S_{m}[g^{ab}_{r},\bar{\phi}_{r}] -\frac{i}{2}
\mbox{tr}\,\, \mbox{ln} \, G_{++}^{0}
-\frac{i}{2} \mbox{tr}\Big[ V_{++}^{(1)}G_{++}^{0}
\nonumber \\ &&
+V_{++}^{(2)}G_{++}^{0}
+\frac{1}{2}V_{++}^{(1)}G_{++}^{0}
V_{++}^{(1)}G_{++}^{0} \nonumber \\
&&+V_{++}^{(1)}G_{+-}^{0}
V_{--}^{(1)}G_{-+}^{0} \Big] \nonumber \\
&&+F_1[g_-^{ab}] + F_2[{\rm det}(g_+)] + O(h^3).
\end{eqnarray}

\subsection{Perturbation expansion for the in-out effective action}
\label{from-inin-to-inout}

Consider now the corresponding calculation in the in-out formalism.
If we compare the generating functionals (\ref{in-out-W}) of the
in-out formalism and (\ref{in-in-W}) of the in-in formalism, we see
that the terms in $\phi$ and in $\phi_+$ in these equations coincide.
Hence, from the definitions (\ref{in-out-effaction}) and
(\ref{in-in-effaction}) of the effective
actions in terms of the generating functionals, it follows that the
in-out propagator and effective action can be obtained from the
corresponding in-in quantities simply by replacing $\phi_+$ with
$\phi$, $J_+$ with $J$, and dropping all terms generated by $\phi_-$
and $J_-$.  The resulting in-out action is
\begin{eqnarray}
\label{in-out-Gamma-approx}
\Gamma_{m}[g^{ab},\bar{\phi}] &=&S_{m}[g^{ab},\bar{\phi}]
-\frac{i}{2} \mbox{tr} \,\, \mbox{ln} \, G^{0}
-\frac{i}{2} \mbox{tr}\Big[ V^{(1)}G^{0} \nonumber \\
&& +V^{(2)}G^{0}
+\frac{1}{2}V^{(1)}G^{0}V^{(1)}G^{0} \Big] +F_2[{\rm det}(g)]\nonumber \\
&& + O(h^3),
\end{eqnarray}
where $G^{0}(x,y) = G(x-y)$ is the usual free Feynman propagator
(\ref{Feynman-prop}), and $V^{(1)}$ and $V^{(2)}$ are the pieces of
the operator $V$ defined in Eq.\ (\ref{V-defn}) that are linear and
quadratic in $h_{ab}$.
Comparing Eq.\ (\ref{in-out-Gamma-approx}) with
the in-in effective action (\ref{Gamma-approx-fin}) and using Eqs.\
(\ref{V-defn}) and (\ref{free-prop-ma}), we see that
the only difference is the term involving $G_{-+}^{0}(x,y)$, since
$G_{++}^0 = G^0$ and $V_{++} = V$.
Hence we can write
\begin{eqnarray}
\label{inout-inin}
\Gamma_{m}[g^{ab}_{+},g^{ab}_{-},\bar{\phi}_{+},\bar{\phi}_{-}]_{{\rm in-in}}
&=&
\Gamma_{m}[g^{ab}_{+},\bar{\phi}_{+}]_{{\rm in-out}} +F_3[g_-^{ab}]
\nonumber \\
&&+U[g^{ab}_{+},g^{ab}_{-},\bar{\phi}_{+},\bar{\phi}_{-}],
\end{eqnarray}
where
\begin{eqnarray}
\label{in-in-part-def}
U[g^{ab}_{+},g^{ab}_{-},\bar{\phi}_{+},\bar{\phi}_{-}]
&=&-\frac{i}{2}\mbox{tr}\left[
V_{++}^{(1)}G_{+-}^{0}V_{--}^{(1)}G_{-+}^{0} \right]
\end{eqnarray}
and $F_3[g_-^{ab}]$ is a term which does not depend on
$g_{+}^{ab}$ or $\bar{\phi}_+$.

\subsection{Explicit calculations}

We write the spacetime metric as
\begin{equation}
g_{ab}=\eta_{ab}+h_{ab},
\end{equation}
where $\eta_{ab}$ is  a flat Minkowski metric.  From now on indices
are raised and lowered with $\eta_{ab}$, and derivatives denoted by a
comma are coordinate derivatives in a Lorentzian coordinate system
associated with the metric $\eta_{ab}$.
Expanding the action (\ref{class-action}) to second order in $h_{ab}$
yields (see Appendix \ref{Notation})
\begin{equation}
\label{action}
S[g^{ab},\phi] = {\dot S}_g[h_{ab}] + {\dot S}_m[h_{ab},\phi] + O(h^3),
\end{equation}
where the quadratic actions ${\dot S}_g$ and ${\dot S}_m$ are given by
\begin{eqnarray}
\label{action-g}
{\dot S}_g[h_{ab}]&&= 2\mu_p^2 \int ({h_{ab}}^{,ab}-h_{,a}^{\ \ a})d^n
x
\end{eqnarray}
and
\FL
\begin{eqnarray}
\label{action-m}
{\dot S}_m[h_{ab},&&\phi] =
-\frac{1}{2}\int(\eta^{ab}\phi_{,a}\phi_{,b}+m^2\phi^2)d^n x
\nonumber \\ &&-\frac{1}{2}\int\Big[
\frac{1}{2}h\eta^{ab}\phi_{,a}\phi_{,b}-h^{ab}\phi_{,a}\phi_{,b}+\frac{1}{2}h
m^2\phi^2\Big]d^n x \nonumber \\
&&-\frac{1}{2}\int\Big[(\frac{1}{8}h^{2}-\frac{1}{4}h^{ab}h_{ab})
\eta^{ab}\phi_{,a}\phi_{,b} \nonumber \\
&&+(\frac{1}{8}h^{2}-\frac{1}{4}h^{ab}h_{ab})m^2\phi^2
-\frac{1}{2}hh^{ab}\phi_{,a}\phi_{,b} \nonumber \\
&&+h^{ac}{h^b}_c\phi_{,a}\phi_{,b}\Big]d^n x.
\end{eqnarray}
Here $h$ is the trace $h \equiv \eta^{ab} h_{ab}$.

Next, we find from Eqs.\ 
(\ref{A-in-in-formula-covariant}) and (\ref{A-in-in-def-1})
the formula for the operator $A_{++}$
\begin{eqnarray}
\label{my-kern}
A_{++}(x,y)&=&  \delta^{n}(x-y) \, (\eta^{ab}\partial_a \partial_b
-m^2 + V),
\end{eqnarray}
where [cf.\ Eq.\ (\ref{V-defn}) above]
\begin{equation}
\label{V}
V = V^{(1)}+V^{(2)}+O(h^3)
\end{equation}
with
\begin{eqnarray}
\label{V1}
V^{(1)}&=&\frac{1}{2} \partial_a h \partial^a - \partial_a
          h^{ab}\partial_b -\frac{1}{2} m^2 h \nonumber \\
          &=&-\partial_a \bar{h}^{ab}\partial_b +\frac{1}{n-2} m^2
          \bar{h}
\end{eqnarray}
and
\FL
\begin{eqnarray}
\label{V2}
V^{(2)}&=&\partial_a \bar{h}^{ac}{\bar{h}^b}_{\ c} \partial_b
  -\frac{1}{4}\partial_a\bar{h}^{cd}\bar{h}_{cd}\eta^{ab}\partial_b
-\frac{1}{n-2}\partial_a \bar{h}\bar{h}^{ab} \partial_b
  \nonumber \\ &&+\frac{1}{4(n-2)}\partial_a \bar{h}^{2} \eta^{ab}
  \partial_b \nonumber \\ &&
+\frac{m^2}{4}\left[\bar{h}^{cd} \bar{h}_{cd}-\frac{1}{n-2}
\bar{h}^2\right].
\end{eqnarray}
Here for simplicity we have written $h_{ab+}$ simply as $h_{ab}$.
Also we are using a notational convention where, for example,
$
(\partial_a h \partial^a) \varphi \equiv \partial_a (h \partial^a
\varphi ).
$
We have also introduced the quantity
\begin{equation}
\bar{h}_{ab}\equiv h_{ab}-\frac{1}{2} h\eta_{ab}
\end{equation}
and its trace $\bar{h}=\eta^{ab}\bar{h}_{ab}$.  Note that ${\bar
h}_{ab}$ is the trace-reversal of the metric perturbation $h_{ab}$
when the dimension $n$ of spacetime is 4, but not otherwise.

Using Eqs.\ (\ref{Gamma-in-in-def}), (\ref{V-defn}),
(\ref{Gamma-approx-fin}), (\ref{V1}), and (\ref{V2}),
we now find that the in-in effective action is
\begin{eqnarray}
\label{ininaction}
\Gamma[h_{ab+},&&h_{ab-},\bar{\phi}_{+},\bar{\phi}_{-}]
={\dot S}[h_{ab+},h_{ab-},\bar{\phi}_{+},\bar{\phi}_{-}]  \nonumber \\
&& +U[h_{ab+},h_{ab-}]
+K_{1}[h_{ab+}]+K_{2}[h_{ab+}]
\nonumber \\ &&
+L[h_{ab+}] + F_4[g_-^{ab},{\rm det}(g_+)] + O(h^3),
\end{eqnarray}
where
\begin{eqnarray}
{\dot S}[h_{ab+},h_{ab-},\bar{\phi}_{+},\bar{\phi}_{-}] && =
{\dot S}_g[h_{ab+}] + {\dot S}_m[h_{ab+},\phi_+]
\nonumber \\ &&
- {\dot S}_g[h_{ab-}] - {\dot S}_m^{*}[h_{ab-},\phi_-].
\end{eqnarray}
In Eq.\ (\ref{ininaction}), we have absorbed the constant
$\mbox{tr}\,\, \ln \, G_{++}^{0}$ and the terms $F_1$ and $F_2$ into
the functional $F_4[g_-^{ab},{\rm det}(g_+)]$.
As before this term is a quantity which
depends on the metric $g_+^{ab}$ only through its determinant, and
which thus affects the in-in equation of motion only via a
renormalization of the cosmological constant.
The terms $U[h_{ab+},h_{ab-}]$, $K_{1}[h_{ab+}]$,
$K_{2}[h_{ab+}]$ and $L[h_{ab+}]$ in Eq.\ (\ref{ininaction}) are given by
\begin{eqnarray}
\label{in-in-part}
U[&h_{ab+}&,h_{ab-}]=-\frac{i}{2}\int d^n x\int d^n x' V^{(1)}[h_{ab+}(x)]
\nonumber \\
    && \ \times G_{+-}^{0}(x-x')\left\{-V^{(1)*}[h_{ab-}(x')] \right\}
G_{-+}^{0}(x'-x) ,
\nonumber \\
\end{eqnarray}
\begin{eqnarray}
\label{K1-def}
K_{1}[h_{ab}]&=& -\frac{i}{2}\int d^n x\int d^n x'\delta^n(x-x')
  V^{(1)}(x)G(x-x')\nonumber \\
&=&-\frac{i}{2}\int d^n x\int d^n x'\delta^n(x-x')
\Big[-\partial_a \bar{h}^{ab}\partial_b  \nonumber \\
&& \ \ \ \ \ \ \ +\frac{1}{n-2} m^2 \bar{h}\Big] G(x-x') ,
\end{eqnarray}
\begin{eqnarray}
K_{2}[h_{ab}]&=& -\frac{i}{2}\int d^n x\int d^n x'\delta^n(x-x')
  V^{(2)}(x)G(x-x')\nonumber \\ &=&-\frac{i}{2}\int d^n x\int d^n
  x'\delta^n(x-x')\Bigg[ \partial_a \bar{h}^{ac}\bar{h}^{b}_{\ c} \partial_b
  \nonumber \\
  &&-\frac{1}{4}\partial_a\bar{h}^{cd}\bar{h}_{cd}\eta^{ab}\partial_b
  -\frac{1}{n-2}\partial_a \bar{h}\bar{h}^{ab} \partial_b \nonumber \\
  &&+\frac{1}{4(n-2)}\partial_a \bar{h}\bar{h}\eta^{ab} \partial_b
  \nonumber \\
  &&+\frac{m^2}{4}(\bar{h}^{cd}\bar{h}_{cd}-\frac{1}{n-2}\bar{h}\bar{h})
  \Bigg] G(x-x'),
\end{eqnarray}
and
\begin{eqnarray}
\label{L-def}
L[h_{ab}]&=&-\frac{i}{4}\int d^n x\int d^n x' \nonumber \\ && \ \ \ \
\times \, V^{(1)}(x) G(x-x') V^{(1)}(x') G(x'-x) \nonumber \\
&=&L_{1}[h_{ab}]+L_{2}[h_{ab}]+L_{3}[h_{ab}]+L_{4}[h_{ab}]  .
\end{eqnarray}
We have also defined
\FL
\begin{eqnarray}
L_{1}[h_{ab}]&=&-\frac{i}{4}\int d^n x \int d^n x' \nonumber \\
         &&\times\bar{h}(x)^{ab}G(x-x')_{,bc'}\bar{h}(x')^{c'd'}G(x'-x)_{,d'a}
         \ ,  \nonumber \\
\end{eqnarray}
\FL
\begin{eqnarray}
L_{2}[h_{ab}]&=&-\frac{i}{4}\int d^n x \int d^n x' \nonumber \\
&&\times\,\frac{m^2}{n-2}\bar{h}(x)G(x-x')_{,c'}\bar{h}(x')^{c'd'}G(x'-x)_{,d'}
\ ,  \nonumber \\
\end{eqnarray}
\FL
\begin{eqnarray}
\label{L3-def}
L_{3}[h_{ab}]&=&-\frac{i}{4}\int d^n x \int d^n x' \nonumber \\
     &&\times \,
     \bar{h}(x)^{ab}G(x-x')_{,b}\frac{m^2}{n-2}\bar{h}(x')G(x'-x)_{,a} \,,
\nonumber \\
\end{eqnarray}
and
\FL
\begin{eqnarray}
\label{L4-def}
L_{4}[h_{ab}]&=&-\frac{i}{4}\int d^n x \int d^n x' \nonumber \\
   &&\times\,\frac{m^2}{n-2}\bar{h}(x) G(x-x') \frac{m^2}{n-2}\bar{h}(x')
   G(x'-x) \ .  \nonumber \\
\end{eqnarray}
In Eqs.\ (\ref{in-in-part}) and (\ref{L-def}), the differential
operators in each factor of $V^{(1)}$ act only on the propagator
immediately to the right of such factors.

It is straightforward to obtain the in-out effective action from the
in-in effective action (\ref{ininaction}) using the method of Sec.\
\ref{from-inin-to-inout} above.  Using Eqs.\ (\ref{inout-inin}) and
(\ref{ininaction}) we obtain
\begin{eqnarray}
\label{inoutaction}
\Gamma[h_{ab},\bar{\phi}]&=&S[h_{ab},\bar{\phi}]
+K_{1}[h_{ab}]+K_{2}[h_{ab}]+L[h_{ab}] \nonumber \\
&& + F_2[{\rm det}(g)] +O(h^3) .
\end{eqnarray}
As already mentioned [cf. Eq.\ (\ref{inout-inin}) above]
the term $U[h_{ab+},h_{ab-}]$ in Eq.\ (\ref{in-in-part}) contains
the differences between the in-in and in-out
formalisms, and has to be added to the in-out effective action
$\Gamma[h_{ab},{\bar \phi}]$ to
obtain the in-in effective action
$\Gamma[h_{ab+},{\bar \phi}_+,h_{ab-},{\bar \phi}_-]$.

Next, we insert the expression (\ref{Feynman-prop}) for the Feynman
propagator $G(x-x^\prime)$ into Eqs.\ (\ref{K1-def})---(\ref{L4-def})
to evaluate the quantities $U$, $K_1$, $K_2$ and $L$.
To simplify the calculations, we work
in the Lorentz gauge where $\bar{h}^{ab}_{\ \ ,b}=0$, and we
regularize the results using dimensional regularization.  The results
are written in terms of curvature invariants using Appendix
\ref{Inv-ito-bar-h}, and are listed in Appendix \ref{KLterms}.

As an example, we show how to compute the term (\ref{L3-def}):
\begin{eqnarray}
L_{3}[&h_{ab}&]=-\frac{i}{4}\int d^n x \int d^n x'
                      \bar{h}^{ab}(x)\frac{m^2}{n-2}\bar{h}(x')
                      \nonumber \\ &&\times\int\frac{d^n
                      p}{(2\pi)^n}\int\frac{d^n q}{(2\pi)^n} \frac{
                      p_b q_a e^{i(p-q)(x-x')}
                      }{(p^2+m^2-i\epsilon)(q^2+m^2-i\epsilon)}
                      \nonumber \\ &=&-\frac{i}{4}\int d^n x \int d^n
                      x'
                      \bar{h}(x)^{ab}\frac{m^2}{n-2}\bar{h}(x')\nonumber
                      \\ &&\times \int\frac{d^n k}{(2\pi)^n}e^{ik(x-x')}
                      \int\frac{d^n q}{(2\pi)^n} \nonumber \\
                      &&\times\frac{q_a
                      (k_b+q_b)}{((k+q)^2+m^2-i\epsilon)(q^2+m^2-i\epsilon)}
                      \ .
\end{eqnarray}
We can drop the $k_b$ in the term $(k_b+q_b)$, since we are working in
the Lorentz gauge.
This yields
\begin{eqnarray}
L_{3}[h_{ab}]&=&-\frac{i}{4}\int d^n x \int d^n x'
             \bar{h}(x)^{ab}\frac{m^2}{n-2}\bar{h}(x') \nonumber \\
             &&\times\int\frac{d^n k}{(2\pi)^n}e^{ik(x-x')} I_{ab}(k)
\end{eqnarray}
where
\begin{eqnarray}
I_{ab}(k)&=&\int\frac{d^n q}{(2\pi)^n} \nonumber \\ &&\times\frac{q_a
    q_b}{((k+q)^2+m^2-i\epsilon)(q^2+m^2-i\epsilon)} \ .
\label{Iab-def}
\end{eqnarray}
In order to perform this integral, we analytically continue to Euclidean
signature.  Defining $I^E_{ab}(k^0,k^j) \equiv I_{ab}(-i k^0,k^j)$ and
using the transformations
\begin{equation}
q^0 \rightarrow -iq^0 \mbox{\ \ \ and \ \ \ } k^0 \rightarrow -ik^0
\
\end{equation}
in Eq.\ (\ref{Iab-def}), we obtain
\begin{eqnarray}
I^{E}_{ab}(k)&=&\int_0^1 dx \int\frac{i d^n q}{(2\pi)^n} \nonumber \\
    &&\times\frac{(i)^{\delta^0_a+\delta^0_b}q_a q_b} {[q^2+m^2+2qk(1-x)+k^2
    (1-x)]^2} \ .
\label{IEab-def}
\end{eqnarray}
In Eq.\ (\ref{IEab-def})  we have also introduced the Feynman
parameter $x$, and dropped
the $i\epsilon$.  The integral over $q$ can now be evaluated (see,
e.g., Ramond \cite{Ramond}). The result is
\begin{eqnarray}
I^{E}_{ab}(k)&=&\int_0^1 dx \frac{i}{(4\pi)^{n/2} \Gamma(2)}\Bigg\{ k_a k_b
    f(k,x) \nonumber \\
    &&+\frac{(i)^{\delta^0_a+\delta^0_b}\delta_{ab}\Gamma(1-n/2)/2}
    {[m^2+k^2 (1-x)-k^2 (1-x)^2]^{1-n/2}} \Bigg\}.
\end{eqnarray}
Next, the term $k_a k_b f(k,x)$ can be dropped, since it will not
contribute to $L_{3}[h_{ab}]$ as we are working in the Lorentz
gauge.  Also, when we analytically continue back to Lorentzian signature
using $k^0 \to i k^0$, we find
$(i)^{\delta^0_a+\delta^0_b}\delta_{ab} \to \eta_{ab}$.
Thus we obtain
\begin{eqnarray}
I_{ab}(k)&=&\frac{i}{(4\pi)^2}\frac{\eta_{ab}}{2-n}
            \Bigg[\frac{2}{4-n}-\gamma+\ln4\pi\Bigg] \nonumber \\
            &&\times\int_0^1 dx \, [m^2+k^2 (1-x)x] \nonumber \\
            &&\times\Big\{1-(2-n/2)\ln[m^2+k^2 (1-x)x]\Big\} \nonumber \\
&&+O[(4-n)] ,
\end{eqnarray}
where $\gamma$ is Euler's constant.
The integral over $x$ can now be performed with the result
\begin{eqnarray}
I_{ab}(k)&=&\frac{i}{(4\pi)^2}\frac{\eta_{ab}}{2-n}\Bigg[
    \Bigg(\frac{2}{4-n}-\gamma-\ln\frac{m^2}{4\pi}+\frac{5}{3}\Bigg)
    \nonumber \\ &&\times\Big(m^2+\frac{k^2}{6}\Big)-\frac{m^2}{3}
    -\frac{1}{3}(k^2+4m^2) \nonumber \\
    &&\times\sqrt{ \frac{k^2+4m^2}{k^2} }
    \mbox{arctanh}\sqrt{ \frac{k^2}{k^2+4m^2} }
    \Bigg]   \nonumber \\
&&+O[(4-n)].
\label{Iab4}
\end{eqnarray}
Below we will write the function
$\mbox{arctanh}(K)=\ln [(1+K)/(1-K)] /2$ appearing in Eq.\ (\ref{Iab4})
in terms of a logarithm. When the $i\epsilon$ from the mass-term
is included this will lead to logarithms of the form
\begin{equation}
\label{def-log}
\ln(K-i\epsilon)\:=\ln|K| -i\pi \Theta(-K) .
\end{equation}
where $\Theta(K)$ is the Heavyside step function. Henceforth when we
write $\ln$ we shall mean the logarithm defined in Eq.\
(\ref{def-log}), which has a branchcut along the negative real axis.

When the remaining terms in Eq.\ (\ref{ininaction}) are evaluated in
the same fashion, and written in terms of curvature invariants
(see Appendix \ref{Inv-ito-bar-h}),
we find the results listed in Appendix \ref{KLterms}. The
effective action (\ref{inoutaction}) then becomes
\begin{equation}
\label{Gamma-ito-W}
\Gamma[h_{ab},\bar{\phi}]=S[h_{ab},\bar{\phi}]+W,
\end{equation}
with
\begin{eqnarray}
\label{W}
W&=&\frac{1}{512\pi^2}\int d^4 x \Bigg[ 8m^4A_{\infty}\sqrt{-g}
     -8\frac{m^2}{3}B_{\infty}\sqrt{-g} R \nonumber \\
     &&+\frac{C_{\infty}}{30}\sqrt{-g} 8R_{ab}R^{ab}
     +\frac{D_{\infty}}{30}\sqrt{-g} 4R^2 \Bigg]   \nonumber \\
&&  + W_{\rm nl}  + F_2[{\rm det}(g)].
\end{eqnarray}
Here
\begin{eqnarray}
\label{Wnl}
W_{\rm nl}&=&\frac{1}{512\pi^2}\int d^4 x\int d^4 x'\int
\frac{d^4 k}{(2\pi)^4} e^{ik(x-x')} \nonumber \\
&&\times\Bigg[R(x)R(x')\tilde{Q}_1(k)
+R_{ab}(x)R^{ab}(x')\tilde{Q}_2(k) \Bigg] \nonumber \\
\end{eqnarray}
is the non-local part of the effective action.
Furthermore we have defined the functions
\begin{eqnarray}
\label{tildeQ1}
\tilde{Q}_1(k)&=&4\Bigg[
-\frac{4}{15}\frac{m^4}{k^4}-\frac{37}{45}\frac{m^2}{k^2}
-\frac{1}{30}\ln\frac{m^2-i\epsilon}{\mu^2} \nonumber \\ &&-\Bigg(
\frac{(k^2+4m^2)^2}{30k^4}
-\frac{2m^2(k^2+4m^2)}{3k^4}+\frac{2m^4}{k^4}\Bigg) \nonumber \\ &&\
\times\sqrt{1+\frac{4(m^2-i\epsilon)}{k^2}} \ln\Bigg(\frac{
\sqrt{1+\frac{4(m^2-i\epsilon)}{k^2}}+1 } {
\sqrt{1+\frac{4(m^2-i\epsilon)}{k^2}}-1 }\Bigg) \Bigg] \nonumber \\
\end{eqnarray}
and
\begin{eqnarray}
\label{tildeQ2}
\tilde{Q}_2(k)=&8\Bigg[&
\frac{16}{15}\frac{m^4}{k^4}+\frac{28}{45}\frac{m^2}{k^2}
-\frac{1}{30}\ln\frac{m^2-i\epsilon}{\mu^2} \nonumber \\
&&-\frac{(k^2+4m^2)^2}{30k^4} \sqrt{1+\frac{4(m^2-i\epsilon)}{k^2}}
\nonumber \\ && \times \ln\Bigg(\frac{
\sqrt{1+\frac{4(m^2-i\epsilon)}{k^2}}+1 } {
\sqrt{1+\frac{4(m^2-i\epsilon)}{k^2}}-1 }\Bigg) \Bigg]  .
\end{eqnarray}
The logarithm used here is the one defined in Eq.\ (\ref{def-log}).
The constants appearing in Eq.\ (\ref{W}) are
\begin{equation}
A_{\infty}=\frac{2}{4-n}+\ln 4\pi-\gamma - \ln \mu^2 +\frac{3}{2}
           -\ln\frac{m^2}{\mu^2}    ,
\end{equation}
\begin{equation}
B_{\infty}=\frac{2}{4-n}+\ln 4\pi-\gamma - \ln \mu^2 +1
            -\ln\frac{m^2}{\mu^2}    ,
\end{equation}
\begin{equation}
\label{C_infty}
C_{\infty}=\frac{2}{4-n}+\ln 4\pi-\gamma - \ln \mu^2 +\frac{46}{15}
\end{equation}
and
\begin{equation}
\label{D_infty}
D_{\infty}=\frac{2}{4-n}+\ln 4\pi-\gamma - \ln \mu^2 +\frac{1}{15}  .
\end{equation}
Note that both $\tilde{Q}_1(k)$ and $\tilde{Q}_2(k)$ are finite at
$k^2 =0$ and that they reduce to
\begin{eqnarray}
\tilde{Q}_2(k)=2\tilde{Q}_1(k) &=& -\frac{4}{15}\ln\Big( \frac{k^2
-i\epsilon}{\mu^2} \Big) \nonumber \\
 &=&-\frac{4}{15}\Big[\ln\Big|\frac{k^2}{\mu^2} \Big|-i\pi\Theta(-k^2)\Big]
\end{eqnarray}
if $m=0$.  Note also that the constant $\mu^2$ appearing in
Eqs.\ (\ref{tildeQ1}) --- (\ref{D_infty})
drops out when these equations are inserted in Eqs.\ (\ref{W}) and
(\ref{Wnl}) in the $m \ne 0$ case.  The  constant $\mu^2$ has
dimension $(\mbox{mass})^2$ and
has been inserted to yield the correct dimensions in the
logarithms of Eqs.\ (\ref{tildeQ1}) and (\ref{tildeQ2}).

Now the functional $F_2[{\rm det}(g)]$ in Eq.\ (\ref{W}) must be a
coordinate invariant, since $\Gamma_m$ and the rest of the terms in
that equation are.  It follows that
\begin{equation}
F_2[{\rm det}(g)] \, \propto \, \int d^n x \, {\sqrt{-g(x)}}.
\label{cosmol1}
\end{equation}

\subsection{Renormalization of the in-out effective action}

We now rewrite
the classical action (\ref{class-action}) in terms of some bare
coupling constants
$\mu_{p_{b}}^2$, $\Lambda_b$, $\alpha_b$ and $\beta_b$:
\begin{eqnarray}
\label{bare-action}
S[g_{ab},\bar{\phi}]&=& -\frac{1}{2}\int d^n x \sqrt{-g}
(\nabla_{a}\bar{\phi}\nabla^{a}\bar{\phi}+m^2\bar{\phi}^2) \nonumber \\
&&+\int d^n x \sqrt{-g} \Bigg[2\mu_{p_{b}}^2 (R-2\Lambda_b ) \nonumber
\\ &&\ \ \ \ -\frac{1}{2}\beta_b R_{ab}R^{ab}-\frac{1}{2}\alpha_b R^2\Bigg]
\end{eqnarray}
{}From Eqs.\ (\ref{Gamma-ito-W}), (\ref{W}) and (\ref{bare-action})
we then find
\begin{eqnarray}
\label{Gamma}
\Gamma[h_{ab},\bar{\phi}]&=& -\frac{1}{2}\int d^4 x \sqrt{-g}
\left[\nabla_{a}\bar{\phi}\nabla^{a}\bar{\phi}+m^2\bar{\phi}^2\right]
\nonumber \\
&&+\int d^4 x \sqrt{-g} \Bigg[2\mu_{p}^2 (R-2\Lambda ) \nonumber \\ &&\
\ \ \ -\frac{1}{2}\beta R_{ab}R^{ab}-\frac{1}{2}\alpha R^2 \Bigg]
\nonumber \\
&& +W_{\rm nl}.
\end{eqnarray}
Here $\mu_{p}^2$, $\Lambda$, $\alpha$ and $\beta$ are
the renormalized values of the parameters,
given by
\begin{eqnarray}
\mu_{p}^2 = \mu_{p_{b}}^2 -\frac{1}{384\pi^2} m^2 B_{\infty} ,
\end{eqnarray}
\begin{eqnarray}
\label{cosmol}
\mu_{p}^2 \Lambda = \mu_{p_{b}}^2 \Lambda_b 
-\frac{1}{256\pi^2}m^4 A_{\infty} + \Delta,
\end{eqnarray}
\begin{eqnarray}
\label{alpha-alpha_b}
\alpha = \alpha_{b} -\frac{1}{1920\pi^2} D_{\infty} ,
\end{eqnarray}
and
\begin{eqnarray}
\label{beta-beta_b}
\beta = \beta_{b} -\frac{1}{960\pi^2} C_{\infty} .
\end{eqnarray}
In the usual way, the renormalized values of the parameters are finite
when we choose the bare parameters suitably.  In Eq.\ (\ref{cosmol}),
the quantity $\Delta$ is the (uncalculated) contribution to the
renormalization of the cosmological constant due to the term
(\ref{cosmol1}).  Note also that the parameters $\alpha$ and $\beta$
which appear in the local part of the renormalized effective action
and the parameter $\mu$ which appears in the non-local part $W_{\rm
nl}$ are not all independent: from Eqs.\ (\ref{Wnl}) -- (\ref{tildeQ2})
and (\ref{Gamma}) it can be seen that a change in $\mu$ can be
compensated for by changes in $\alpha$ and $\beta$.

Finally, the renormalized in-in effective action is given by combining
Eqs.\ (\ref{inout-inin}), (\ref{Gamma}), and Eq.\ (\ref{U-formula}) from
Appendix \ref{KLterms} below.

\section{The stress energy tensor}
\label{stress-energy}

\subsection{Equations of motion in the in-out formalism}

The semiclassical equations of motion are obtained from
\begin{equation}
\label{eq-of-motion}
\frac{-2}{\sqrt{-g}}\frac{\delta}{\delta g^{ab}}\Gamma[h_{cd},\bar{\phi}]=0,
\end{equation}
and
\begin{equation}
\label{eq-of-motion-1}
\frac{\delta}{\delta {\bar \phi}}\Gamma[h_{cd},\bar{\phi}]=0.
\end{equation}
Equation (\ref{eq-of-motion-1}) is automatically solved when we choose
\begin{equation}
{\bar \phi}  = {\bar \phi}[g^{ab}],
\label{vp}
\end{equation}
the functional given by Eq.\ (\ref{def-phi-bar}) above at $J=0$.
When we insert Eqs.\ (\ref{Gamma}) and (\ref{vp}) into Eq.\
(\ref{eq-of-motion}), the functional derivative of the first term on
the right hand side of Eq.\ (\ref{Gamma}) can be dropped since it is
of order $O(h^2)$ [as ${\bar \phi}$ from Eq.\ (\ref{def-phi-bar})
for $J=0$ is of order $O(h)$].  The resulting equation of motion is
\begin{eqnarray}
\label{motion}
G_{ab}(x)+\Lambda g_{ab}(x)
&=&\frac{1}{4\mu_p^2} T_{ab}(x)_{{\rm in-out}}  ,
\end{eqnarray}
where we have defined the in-out expected stress-energy tensor
\begin{eqnarray}
\label{Tab-in-out}
T_{ab}(x)_{{\rm in-out}}& \equiv &
\frac{\langle 0,\mbox{out}|\hat{T}_{ab}|0,\mbox{in} \rangle}
     {\langle 0,\mbox{out}|0,\mbox{in} \rangle}
\nonumber \\
&=&\frac{-1}{256\pi^2}\int d^4 x'
   \int \frac{d^4 k}{(2\pi)^4} e^{ik(x-x')} \nonumber \\
      &&\times\Bigg[\dot{H}^{(1)}_{ab}(x')\tilde{Q}_1(k)
                   +\dot{H}^{(2)}_{ab}(x')\tilde{Q}_2(k)
              \Bigg]   \nonumber \\
&&+\alpha \dot{H}^{(1)}_{ab}(x)+ \beta \dot{H}^{(2)}_{ab}(x)  .
\end{eqnarray}
In writing this tensor we have also introduced
the linearized versions $\dot{H}^{(1)}_{ab}(x)$
and $\dot{H}^{(2)}_{ab}(x)$ [see Eqs.\ (\ref{dotH1}) and (\ref{dotH2}) below]
of the conserved local curvature tensors
\begin{eqnarray}
\label{H1}
H^{(1)}_{ab}(x)=\frac{1}{\sqrt{-g}}\frac{\delta}{\delta g^{ab}(x)}
\int d^4 x' \sqrt{-g} R(x')R(x')   \nonumber \\
=2g_{ab}\Box R - 2\nabla_{a}\nabla_{b}R
 -\frac{1}{2}g_{ab}R^2 + 2RR_{ab}
\end{eqnarray}
and
\begin{eqnarray}
\label{H2}
H^{(2)}_{ab}(x)&=&\frac{1}{\sqrt{-g}}\frac{\delta}{\delta g^{ab}(x)}
\int d^4 x' \sqrt{-g} R_{cd}(x')R^{cd}(x')    \nonumber \\
&=&\frac{1}{2}g_{ab}\Box R +\Box R_{ab}
-2\nabla_{c}\nabla_{a} R^{c}_{\ b}  \nonumber \\
&&+2R^{c}_{\ a}R_{cb}  -\frac{1}{2}g_{ab}R_{cd}R^{cd}  .
\end{eqnarray}

As is well known, Eq.\ (\ref{motion}) is not a physically realistic
equation for semiclassical gravity since the right hand side is
complex and not real \cite{J1,J2}.

\subsection{The in-in expected stress-energy tensor}

By combining Eqs.\ (\ref{bar-phi-value}), (\ref{Gamma-eqn-of-motion-1}),
(\ref{ininaction}), (\ref{inoutaction}),
(\ref{eq-of-motion}) and (\ref{motion})
we obtain the equations of motion in the in-in formalism
\begin{eqnarray}
\label{ininmotion}
G_{ab}(x)+\Lambda g_{ab}(x)
&=&\frac{1}{4\mu_p^2} \big[T_{ab}(x)_{{\rm in-out}} +T'_{ab}(x)\big] ,
\end{eqnarray}
where $T_{ab}(x)_{{\rm in-out}}$ is given by Eq.\ (\ref{Tab-in-out}),
and where the additional term $T'_{ab}(x)$ due to the term $U$
in Eq.\ (\ref{ininaction}) is given by
\begin{eqnarray}
\label{T'ab}
{T'}_{ab}(x)&=&
\frac{-2}{\sqrt{-g}}\frac{\delta U[h_{ab+},h_{ab-}]}
       {\delta g^{ab}_{+}}\Big|_{g^{ab}_{+}=g^{ab}_{-}=g_{ab}}\nonumber \\
&=&\frac{-1}{256\pi^2}\int d^4 x'\int \frac{d^4 k}{(2\pi)^4}
e^{ik(x-x')}  \nonumber \\
       &&\times \Bigg[\dot{H}^{(1)}_{ab}(x')\tilde{Q}_1'(k)
                   +\dot{H}^{(2)}_{ab}(x')\tilde{Q}_2'(k) \Bigg].
\end{eqnarray}
Here we used Eq.\ (\ref{U-formula}) from Appendix \ref{KLterms} below,
and have defined
\begin{eqnarray}
\label{tildeQ1'}
\tilde{Q}_1'(k)&=&-4\Bigg[
 \frac{(k^2+4m^2)^2}{30k^4}
-\frac{2m^2(k^2+4m^2)}{3k^4}+\frac{2m^4}{k^4}\Bigg] \nonumber \\
   &&\times \sqrt{1+\frac{4m^2}{k^2}}
     2\pi i \Theta(-k^2 -4m^2)\Theta(-k^0)
\end{eqnarray}
and
\begin{eqnarray}
\label{tildeQ2'}
\tilde{Q}_2'(k)&=&-8\Bigg[\frac{(k^2+4m^2)^2}{30k^4}\Bigg] \nonumber \\
 &&\times \sqrt{1+\frac{4m^2}{k^2}}
    2\pi i \Theta(-k^2 -4m^2)\Theta(-k^0).
\end{eqnarray}

The in-in expected stress-energy tensor
\begin{eqnarray}
T_{ab}(x)_{{\rm in-in}} \equiv
\frac{\langle 0,\mbox{in}|\hat{T}_{ab}|0,\mbox{in} \rangle}
     {\langle 0,\mbox{in}|0,\mbox{in} \rangle}
\end{eqnarray}
is therefore given by
\begin{eqnarray}
\label{Tab-in-in}
T_{ab}(x)_{{\rm in-in}}&=&T_{ab}(x)_{{\rm in-out}}+{T'}_{ab}(x)
\nonumber \\
&=&\frac{-1}{256\pi^2}\int d^4 x'\int \frac{d^4 k}{(2\pi)^4}
e^{ik(x-x')}  \nonumber \\
      &&\times\Bigg[\dot{H}^{(1)}_{ab}(x')\tilde{T}_1(k)
                   +\dot{H}^{(2)}_{ab}(x')\tilde{T}_2(k)
              \Bigg]   \nonumber \\
&&+\alpha \dot{H}^{(1)}_{ab}(x)+ \beta \dot{H}^{(2)}_{ab}(x)  .
\end{eqnarray}
Here we have defined
\begin{eqnarray}
\label{tildeT1}
\tilde{T}_1(k)&=&\tilde{Q}_1(k)+\tilde{Q}_1'(k)    \nonumber \\
&=&4\Bigg\{
-\frac{4}{15}\frac{m^4}{k^4}-\frac{37}{45}\frac{m^2}{k^2}
-\frac{1}{30}\ln\frac{m^2-i\epsilon}{\mu^2} \nonumber \\ &&-\Bigg[
\frac{(k^2+4m^2)^2}{30k^4}
-\frac{2m^2(k^2+4m^2)}{3k^4}+\frac{2m^4}{k^4}\Bigg] \nonumber \\
&&\times\sqrt{1+\frac{4(m^2-i\epsilon)}{k^2}}      \nonumber \\
&&\times\ln\Bigg(\frac{
\sqrt{1+\frac{4(m^2-i\epsilon{\rm Sgn}k^0)}{k^2}}+1 } {
\sqrt{1+\frac{4(m^2-i\epsilon{\rm Sgn}k^0)}{k^2}}-1 }\Bigg) \Bigg\}
\nonumber \\
\end{eqnarray}
and
\begin{eqnarray}
\label{tildeT2}
\tilde{T}_2(k)=&&\tilde{Q}_2(k)+\tilde{Q}_2'(k)    \nonumber \\
=&8\Bigg[&
\frac{16}{15}\frac{m^4}{k^4}+\frac{28}{45}\frac{m^2}{k^2}
-\frac{1}{30}\ln\frac{m^2-i\epsilon}{\mu^2} \nonumber \\
&&-\frac{(k^2+4m^2)^2}{30k^4} \sqrt{1+\frac{4(m^2-i\epsilon)}{k^2}}
\nonumber \\
&& \times\ln\Bigg(\frac{
\sqrt{1+\frac{4(m^2-i\epsilon{\rm Sgn}k^0)}{k^2}}+1 } {
\sqrt{1+\frac{4(m^2-i\epsilon{\rm Sgn}k^0)}{k^2}}-1 }\Bigg) \Bigg]  .
\end{eqnarray}
It is easy to see that ${\tilde T}_1(k)$ and ${\tilde T}_2(k)$ are
finite at $k^2=0$ (for $m \ne 0$), and that they are sufficiently
regular that their Fourier transforms $T_1(x)$ and $T_2(x)$ exist as
distributions.

\section{Properties of the in-in expected stress-energy tensor}
\label{Properties}

The stress-energy tensor given in Eq.\ (\ref{Tab-in-in}) is
determined by the Green functions $\tilde{T}_1(k)$ and $\tilde{T}_2(k)$
in Eqs.\ (\ref{tildeT1}) and (\ref{tildeT2}).
In this section we show that in the limit $m \to 0$,
these Green functions reduce
to the previously obtained Green functions for a massless
field.  We also show that they are causal, i.e., that their Fourier
transforms $T_1(x)$ and $T_2(x)$ have
support only inside the past light cone.  These properties serve as a
check of our calculation.

\subsection{The massless limit}
\label{massless-limit}
The Green functions $\tilde{T}_1(k)$ and $\tilde{T}_2(k)$
in Eqs.\ (\ref{tildeT1}) and (\ref{tildeT2}) reduce to
\begin{eqnarray}
\tilde{T}_2(k)=2\tilde{T}_1(k) &=&-\frac{4}{15}\ln\Big( \frac{k^2
-i\epsilon {\rm Sgn}(k^{0})}{\mu^2} \Big)  \nonumber \\
&=&-\frac{4}{15}\Bigg[\ln\Big( \frac{|k^2|}{\mu^2} \Big)
-i\pi\Theta(-k^{2}){\rm Sgn}(k^{0}) \Bigg]  \nonumber \\
\end{eqnarray}
if $m=0$.  This Green function together with
Eq.\ (\ref{Tab-in-in}) yields exactly the same
the stress-energy tensor as found by Horowitz \cite{Horowitz}
and Jordan \cite{J2}.

Note that the Green functions are not smooth in $m^2$ near
$m =0$.  For $k^0 > 0$ we find
\begin{eqnarray}
\frac{\partial\tilde{T}_{2}}{\partial m^2}\Big|_{m^2=0}
&=& 8\Bigg[\frac{28}{45k^2}-
\Bigg(\frac{1}{15k^2\sqrt{1-\frac{4i\epsilon}{k^2}}}
+\frac{4\sqrt{1-\frac{4i\epsilon}{k^2}}}{15k^2}\Bigg) \nonumber \\
&&\times\ln\Bigg(\frac{
\sqrt{1-\frac{4i\epsilon}{k^2}}+1 } {
\sqrt{1-\frac{4i\epsilon}{k^2}}-1 }\Bigg) \Bigg],
\end{eqnarray}
which diverges in the limit $\epsilon \rightarrow 0$.  Hence the first
derivative of the stress tensor with respect to $m^2$ does not exist
at $m=0$.

\subsection{Causality}

It is difficult to find the Fourier transforms $T_{j}(x)$, $j = 1,2$,
of the Green functions (\ref{tildeT1}) and (\ref{tildeT2}).  However,
it is not necessary to explicitly perform these Fourier transforms in
order to demonstrate causality.  By Lorentz invariance it is
sufficient to show that
\begin{eqnarray}
T_{j}(t,0,0,0)= 0   \ \ \  \mbox{ for \ } t<0
\label{causality-1}
\end{eqnarray}
and
\begin{eqnarray}
T_{j}(0,r,0,0)= 0   \ \ \  \mbox{ for \ } r\neq 0,
\label{causality-2}
\end{eqnarray}
for $j = 1,2$.  In other words, the Green functions $T_{j}(x)$  must
be zero inside the past light cone and outside the light cone.
To check the condition (\ref{causality-1}) we write
\begin{eqnarray}
T_{j}(t,0,0,0)= \frac{1}{(2\pi)^4}\int d^3 k \int d k^0
e^{-i k^0 t} \tilde{T}_{j}(k)  .
\end{eqnarray}
Now from Eqs.\ (\ref{tildeT1}) and (\ref{tildeT2}) we
see that the logarithmic terms in both ${\tilde T}_1(k)$ and ${\tilde
T}_2(k)$ have branchcuts
in the lower complex $k^0$ plane, but no poles elsewhere.
It is therefore possible
to deform the contour of the $k^0$ integration
into the usual semi-circle with infinite radius
in the upper complex $k^0$ plane. Since $t<0$ the integral vanishes.
This immediately shows that $T_{j}(t,0,0,0)=0$ for $t<0$.
A similar argument can be used to show that
$T_{j}(0,r,0,0)= 0$.

\section{Conclusions}
\label{Conclusion}

We have shown that the Wald axioms determine the stress-energy tensor
(up to two parameters) only in the case of a massless field. In the
case of a massive scalar field, the Wald axioms allow for a much
larger ambiguity.  We have calculated the expectation value of the
stress-energy
tensor in the incoming vacuum state for a massive scalar field on any
spacetime which is a linear perturbation off Minkowski spacetime,
generalizing an earlier formula of Horowitz \cite{Horowitz} and Jordan
\cite{J2} in the massless case.  In our calculation we used the in-in
effective action formalism \cite{Schwing,J1,CampVar}.  As expected,
the resulting stress-energy tensor is causal and reduces to the known
result in the massless case in the limit $m \rightarrow 0$.  As in the
massless case, we find a two parameter ambiguity in the stress-energy
tensor, even though this is not guaranteed by the Wald axioms.

After this paper was submitted for publication, we became aware of
related work by Dalvit and Mazzitelli \cite{Dalvit}.  Dalvit and
Mazzitelli calculate an in-in effective action using the method of
expanding in powers of curvature pioneered by Vilkovisky and
collaborators \cite{Vilkovisky}.  Our result (\ref{intro-Tab-in-in})
can be derived from Eqs.\ (38) and (41) of Ref.\ \cite{Dalvit} by
evaluating the integrals over $t$, by specializing to linear
perturbations about flat spacetime, and by representating the
d'Alembertian operators in Fourier space.  Our derivation of the
result (\ref{intro-Tab-in-in}) has the advantage that it is more direct
and simple than that of Ref.\ \cite{Dalvit}, since it does not rely
on the Vilkovisky expansion formalism.

We conclude by listing some open questions.  First, are there
additional axioms which would reduce the ambiguity in the
stress-energy tensor?  Second, will the same stress-energy tensor
(\ref{intro-Tab-in-in}) be
predicted (up to the two parameter ambiguity) by the point splitting
method \cite{WaldQFT} or by the deWitt-Schwinger method \cite{BirDav}?
There is no a-priori guarantee that this will be the case.

\acknowledgments
We thank Robert Wald for helpful conversations.  This research was
supported in part by NSF grant PHY--9722189 and by a Sloan Foundation
fellowship.

\appendix

\section{Notation and conventions}
\label{Notation}

We use units in which $\hbar=c=1$, but in which $G\neq 1$, so that the
Planck mass
is given by
\begin{equation}
\mu_p=\sqrt{\frac{1}{32\pi G}}.
\end{equation}
Throughout we use the same sign conventions
for metric and curvature tensors as in the book of Misner, Thorne, and
Wheeler \cite{MTW}.  Specifically the metric $g_{ab}$ has signature
$(-,+,+,+)$.  Indices $i,j,k, \ldots$ run over the spatial indices
$1,2,3$ while indices $a,b,c, \ldots$ run over $0,1,2,3$.

We introduce the metric perturbation
\begin{equation}
h_{ab}=g_{ab}-\eta_{ab}
\end{equation}
and its trace $h =h_{ab}\eta^{ab}$, where $\eta_{ab}$ is a flat metric.
In expressions involving $h_{ab}$, indices are raised and lowered with
the flat spacetime metric $\eta^{ab}$.
The coordinate derivative of a tensor $T^{a}_{\ b}$
in a Lorentzian coordinate system with respect to the metric $\eta_{ab}$
is denoted in the usual way:
\begin{equation}
T^{a}_{\ b,c}=\partial_c T^{a}_{\ b}  .
\end{equation}
The Fourier transform of any function $F(x)$ on Minkowski spacetime is
defined as
\begin{eqnarray}
\tilde{F}(k)=\int d^4 x \ e^{-ik x} F(x).
\end{eqnarray}
Throughout we use
\begin{equation}
kx=\eta_{ab}k^a x^b
\end{equation}
to denote the dot product of two 4-vectors $k^a$ and $x^a$ in
Minkowski spacetime.  We use $\Theta(x)$ to denote the step function,
$\Theta(x) = 1$ for $x>0$ and $0$ otherwise.

Products of operators $A$,$B$ are defined as
\begin{equation}
(AB)(x,z) = \int d^4 y \, A(x,y) \, B(y,z).
\end{equation}
Thus, factors of $\sqrt{-g}$ are not implicit in expressions such
as $G V$ in our calculations.  We show such factors explicitly when
they are required, with the exception of the notation for products of
functions
\begin{equation}
\langle f_1,f_2 \rangle  \equiv \int d^n x \, \sqrt{-g(x)} f_1(x)^* f_2(x).
\end{equation}

\section{Expressions for curvature invariants}
\label{Inv-ito-bar-h}

In this appendix we expand the various possible local counterterms in
the effective action to second order in the metric perturbation, and
write them in terms of the quantity ${\bar h}_{ab}$.  We define
\begin{equation}
\bar{h}_{ab}\:= h_{ab}-\frac{1}{2}h \eta_{ab},
\end{equation}
where $h=\eta^{ab}h_{ab}$.  In $n$ dimensions we then find
\begin{equation}
h=-\frac{2}{n-2}\bar{h}
\end{equation}
and thus
\begin{equation}
h_{ab}=\bar{h}_{ab}-\frac{1}{n-2}\bar{h} \eta_{ab}.
\end{equation}
In our calculation of the effective action, when we expand quantities
such as $\sqrt{-g}$, $R$ or $R_{ab}$ in terms of
$\bar{h}_{ab}$ and $\bar{h}$, we find terms of order $O(n-4)$.  Such
terms must sometimes be kept and not discarded, since they can give rise
to finite contributions when multiplied by infinite terms
of the form
$1/(n-4)$.

We find for the determinant of the metric tensor
\begin{equation}
\sqrt{-g} = 1-\frac{1}{n-2}\bar{h}-\frac{1}{4}\bar{h}^{cd}\bar{h}_{cd}
          +\frac{1}{4(n-2)}\bar{h}\bar{h} +O(h^3).
\end{equation}
The curvature scalar becomes
\begin{equation}
R=R^{(1)}+R^{(2)}+O(h^3) ,
\end{equation}
where
\begin{equation}
R^{(1)}=\frac{1}{n-2}\bar{h}_{,a}^{\ \ a} +\bar{h}_{ab}^{\ \ ,ab}
\end{equation}
and
\begin{eqnarray}
\label{curv2}
R^{(2)}&=&\frac{3}{4}\bar{h}_{ab,c}\bar{h}^{ab,c}
+\bar{h}_{ab}\bar{h}^{ab\ \ c}_{\ \ ,c}
-\frac{3n-10}{4(n-2)^2}\bar{h}_{,c}\bar{h}^{,c}  \nonumber \\
&&-\frac{(n-4)}{(n-2)^2}\bar{h}_{,c}^{\ \ c}\bar{h}
-\frac{1}{2}\bar{h}^{ca,b}\bar{h}_{cb,a}         \nonumber \\
&&+\frac{1}{n-2}\Big(2\bar{h}\bar{h}_{ab}^{\ \ ,ab}
+\bar{h}_{,a}\bar{h}^{ab}_{\ \
,b}
\Big)       \nonumber \\ &&-2\bar{h}^{ab}\bar{h}_{ac,b}^{\ \ \ \ c}
-\bar{h}^{ab}_{\ \ ,b}\bar{h}_{ac}^{\ \ ,c} .
\end{eqnarray}
Other useful scalars are
\begin{eqnarray}
R^2&=&\frac{1}{(n-2)^2}\bar{h}_{,a}^{\ \ a}\bar{h}_{,b}^{\ \ b} \nonumber \\
&&+\Big(\frac{2}{n-2}\bar{h}_{,c}^{\ \ c}
  +\bar{h}_{cd}^{\ \ ,cd}\Big)\bar{h}_{ab}^{\ \ ,ab} + O(h^3)
\end{eqnarray}
and
\begin{eqnarray}
R_{ab}R^{ab}&=&\frac{1}{4}\left[
 \bar{h}_{ab\ \ c}^{\ \ ,c}\bar{h}^{ab\ \ d}_{\ \ ,d}
-\frac{(n-4)}{(n-2)^2}\bar{h}_{,a}^{\ \ a}\bar{h}_{,b}^{\ \ b}
                                                          \right] \nonumber \\
&+&\frac{1}{4}\Big(-2\bar{h}_{ab,c}^{\ \ \ \ c}
    +\frac{2}{n-2}\bar{h}_{,c}^{\ \ c}\eta_{ab}
    +\bar{h}_{ca,b}^{\ \ \ \ c}+\bar{h}_{cb,a}^{\ \ \ \ c}\Big) \nonumber \\
&&\times \Big(\bar{h}_{c}^{\ a,bc}+\bar{h}_{c}^{\ b,ac}\Big) +O(h^3) .
\end{eqnarray}
Note that many of the terms vanish if the  Lorentz gauge
$\bar{h}^{ab}_{\ \ ,b}=0$ is used.
In the Lorentz gauge we find that
(discarding surface terms)
\begin{eqnarray}
\int d^n x 8\sqrt{-g} &&=
 \int d^4 x(8-4\bar{h}-2\bar{h}_{ab}\bar{h}^{ab}+\bar{h}\bar{h}) \nonumber \\
&&+ {1 \over 2} \delta
\int d^4 x (4\bar{h}-\bar{h}\bar{h})+O(\delta^2),
\end{eqnarray}
\begin{eqnarray}
\int d^n x 8\sqrt{-g} R &=&
\int d^4 x(2{\bar{h}_{ab,c}}^{\ \ \ \ c}\bar{h}^{ab}
                          -\bar{h}_{,c}^{\ \ c}\bar{h})   \nonumber \\
&&+ {1\over2} \delta \int d^4 x \bar{h}_{,c}^{\ \ c}\bar{h} +O(\delta^2),
\end{eqnarray}
\begin{eqnarray}
\int d^n x 4\sqrt{-g} R^2 &=&
\int d^4 x\bar{h}_{,a}^{\ \ a}\bar{h}_{,b}^{\ \ b}      \nonumber \\
&&- {1\over 2} \delta
\int d^4 x 2\bar{h}_{,a}^{\ \ a}\bar{h}_{,b}^{\ \ b}
+ O(\delta^2),
\end{eqnarray}
and
\begin{eqnarray}
\int d^n x 8\sqrt{-g} R^{ab}R_{ab} &=&
\int d^4 x 2{\bar{h}^{ab\ c}}_{\ \ ,c}{\bar{h}_{ab,d}}^{\ \ \ \ d}
\nonumber \\
&& - {1 \over 2} \delta
\int d^4 x \bar{h}_{,c}^{\ \ c}\bar{h}_{,d}^{\ \ d} \nonumber \\
&&+ O(\delta^2),
\end{eqnarray}
where $\delta = n - 4$.

Using the Lorentz gauge and specializing to four dimensions,
we find for the linearized versions of the local curvature
tensors (\ref{H1}) and (\ref{H2})
\begin{eqnarray}
\label{dotH1}
\dot{H}^{(1)}_{ab}(x)=
\eta_{ab}\bar{h}_{,c\ d}^{\ \ c\ d}-\bar{h}_{,abc}^{\ \ \ \ c}
+O(h^2)
\end{eqnarray}
and
\begin{eqnarray}
\label{dotH2}
\dot{H}^{(2)}_{ab}(x)&=&
\frac{1}{2}\Big(\eta_{ab}\bar{h}_{,c\ d}^{\ \ c\ d}
 -\bar{h}_{,abc}^{\ \ \ \ c}- \bar{h}_{ab,c\ d}^{\ \ \ \ c\ d}
\Big) \nonumber \\
&&+O(h^2).
\end{eqnarray}

\section{The in-in propagator in Minkowski spacetime}
\label{App-invers}

In this Appendix we combine Eqs.\ (\ref{A-in-in-formula-covariant}),
(\ref{A-in-in-def-1}) and (\ref{prop-def}) specialized to Minkowski
spacetime to obtain Eqs.\ (\ref{free-prop-ma})-(\ref{wighty}).  Using
Eq.\ (\ref{prop-def}) 
and the expansions (\ref{A-expansion}) and (\ref{G-expansion}), we
obtain
\begin{eqnarray}
\label{free-prop-def}
\int d^n x' A_{rs}^{0}(x,x') G_{st}^{0}(x',y) = -\delta^n (x-y) \delta_{rt}.
\end{eqnarray}
Note that it follows from the form of $A_{rs}(x,y)$ given by 
Eqs.\ (\ref{A-in-in-formula-covariant}) and (\ref{A-in-in-def-1}) that
$G_{--}^{0}(x,y)=-G_{++}^{0*}(x,y)$ 
and that $G_{-+}^{0}(x,y)=-G_{+-}^{0*}(x,y)$.

First, the relation
\begin{equation}
G_{++}^{0}(x,y) = G(x-y),
\label{G++-value}
\end{equation}
follows immediately from translational invariance and the equation
\begin{eqnarray}
[\Box_{x} -(m^2-i\epsilon)] G_{++}^{0}(x,y) = -\delta^{n} (x-y)
\label{c3e}
\end{eqnarray}
which follows from Eq.\ (\ref{free-prop-def}).

The equation determining $G_{-+}^{0}(x,y)$ is, from Eq.\
(\ref{free-prop-def}),
\begin{eqnarray}
[\Box_{x} -(m^2+i\epsilon)] G_{-+}^{0}(x,y) = 0 ,
\end{eqnarray}
which upon Fourier transforming and using translational invariance
becomes
\begin{eqnarray}
\label{eq-Gp}
[-p^2 -(m^2+i\epsilon)] \tilde{G}_{-+}^{0}(p) = 0 .
\end{eqnarray}
Any function $\tilde{G}_{-+}^{0}(p)$ which has support only on
the hypersurface $p^2=-(m^2+i\epsilon)$
will be a solution to this equation, and
hence $\tilde{G}_{-+}^{0}(p)$ is not uniquely determined by
Eq.\ (\ref{eq-Gp}).
What we have not used yet is the boundary condition that
$\phi_{+}=\phi_{-}$ on the hypersurface given by
$x^0=T$, where $T\rightarrow \infty$.
To make use of this boundary condition
note that the classical equations for $\phi_{\pm}$ are
\begin{eqnarray}
\int d^n y \, A_{rs}^0(x,y) \phi_{s}(y) = -{\hat J}_{r}(x),
\end{eqnarray}
and that the solutions to these equations are
\begin{eqnarray}
\phi_{s}(x) =  \int d^n y \, G_{st}^{0}(x,y) {\hat J}_{t}(y).
\end{eqnarray}
Enforcing the above mentioned boundary condition yields
\begin{eqnarray}
\int d^n y \, G_{+s}^{0}(x,y) {\hat J}_{s}(y) =\int d^n y \,
G_{-s}^{0}(x,y) {\hat J}_{s}(y)
\end{eqnarray}
at $x = (T, x^1,x^2,x^3)$, which using Eq.\ (\ref{hatJ-def}) simplifies to
\begin{eqnarray}
&\int d^n y \left[G_{++}^{0}(x,y)-G_{-+}^{0}(x,y)\right] J_{+}(y)&
\nonumber \\
+&\int d^n y \left[G_{+-}^{0}(x,y)-G_{--}^{0}(x,y)\right] J_{-}(y)&  =
0.
\label{c2c}
\end{eqnarray}
Since the sources $J_{\pm}(y)$ are completely arbitrary, Eqs.\
(\ref{G++-value}) and (\ref{c2c}) imply that
\begin{eqnarray}
\label{rel-G-G_pm}
G_{-+}^{0}(x,y)=G(x-y)
\end{eqnarray}
for $x=(T, x^1,x^2,x^3)$.  Now the
Feynman propagator $G(x)$ can be written as
\begin{equation}
\label{rel-G-wighty}
G(x)=-\Theta(x^0)\Delta^{-}(x) +\Theta(-x^0)\Delta^{+}(x),
\end{equation}
where
\FL
\begin{equation}
\Delta^{\pm}(x)= \pm 2\pi i\int \frac{d^n p}{(2\pi)^n}
e^{ip(x-y)} \delta (p^2+m^2)\Theta{(\mp p^0)}
\end{equation}
are the positive and negative Wightman functions \cite{Peskin}.
{}From Eqs.\ (\ref{rel-G-G_pm}) and (\ref{rel-G-wighty}) it follows that
\begin{eqnarray}
G_{-+}^{0}(x,y)=-\Delta^{-}(x-y)
\label{anss}
\end{eqnarray}
for large $x^0$.  Equation (\ref{anss}) now implies that
the appropriate solution of Eq.\ (\ref{eq-Gp}) which fulfills the
boundary condition at $x^0=T$ corresponds to $G^0_{-+}(x,y) = -
\Delta^{-}(x-y)$, which yields Eq.\ (\ref{free-prop-ma}) above.

\section{Expansion of the propagator}
\label{expandprop}

In this Appendix we obtain the expansion (\ref{log-G++}) for the
operator $G_{++}$.
{}From Eqs.\ (\ref{prop-def}), (\ref{A-expansion}), (\ref{G-expansion})
and (\ref{free-prop-def}) it follows that
\begin{eqnarray}
G_{rt} &=& G_{rs}^{0} \left[\delta_{st}+V_{ss'}G_{s't}^{0}
+V_{ss'}G_{s's''}^{0}V_{s''t'}G_{t't}^{0} \right]
\nonumber \\ &&
+ O(V^3).
\end{eqnarray}
This implies that
\begin{eqnarray}
G_{++}&=&G_{++}^{0} +G_{+s}^{0}V_{ss'}G_{s'+}^{0} \nonumber \\
&&+G_{+s}^{0}V_{ss'}G_{s's''}^{0}V_{s''t}G_{t+}^{0} + ...  \nonumber \\
&=&G_{++}^{0}\big[1+V_{+s}G_{s+}^{0}+V_{+s}G_{ss'}^{0}V_{s't}G_{t+}^{0}
+ O(V^3) \big]\ ,  \nonumber \\
\end{eqnarray}
where we have used the fact that
$G_{++}^{0^{-1}}G_{+-}^{0}=A_{++}^{0}G_{+-}^{0}=0$.
Hence we can write the logarithm of the propagator as
\begin{eqnarray}
\label{log-G1}
\mbox{ln}(G_{++}) &=&\mbox{ln}(G_{++}^{0})
+V_{+s}G_{s+}^{0}+V_{+s}G_{ss'}^{0}V_{s't}G_{t+}^{0}     \nonumber \\
&&-\frac{1}{2}V_{+s}G_{s+}^{0}V_{+t}G_{t+}^{0} + O(V^3).
\end{eqnarray}
Next we use the fact that $V_{rs}$ is diagonal [cf.\ Eq.\
(\ref{V-defn}) above] to obtain
\begin{eqnarray}
\mbox{ln}(G_{++}) &=&\mbox{ln}(G_{++}^{0})+V_{++}G_{++}^{0}
+\frac{1}{2}V_{++}G_{++}^{0}V_{++}G_{++}^{0} \nonumber \\
&&+V_{++}G_{+-}^{0}V_{--}G_{-+}^{0}+ O(V^3)  \nonumber \\
&=&\mbox{ln}(G_{++}^{0}) +V_{++}^{(1)}G_{++}^{0}
+V_{++}^{(2)}G_{++}^{0}         \nonumber \\
&&+\frac{1}{2}V_{++}^{(1)}G_{++}^{0}V_{++}^{(1)}G_{++}^{0}
+V_{++}^{(1)}G_{+-}^{0}V_{--}^{(1)}G_{-+}^{0}   \nonumber \\
&&+ O(V^3)  \ ,
\end{eqnarray}
which yields Eq.\ (\ref{log-G++}).

\section{Terms in the effective action}
\label{KLterms}

In this Appendix we list the terms contributing to the effective
action (\ref{ininaction})
given by Eqs.\ (\ref{in-in-part})-(\ref{L4-def}). Throughout we use the
Lorentz gauge.  We find
\begin{eqnarray}
K_{1}[h_{ab}]&=&\frac{1}{512\pi^2}\int d^4 x \bar{h}(x)
(-4m^2)\Big(Y+\frac{5}{2}\Big) ,
\end{eqnarray}
\begin{eqnarray}
K_{2}[h_{ab}]&=&\frac{1}{512\pi^2}\int d^4 x \Bigg[
\bar{h}(x)\bar{h}(x)\Big( 2m^4 \Big)
\Big(Y+\frac{5}{2}\Big)   \nonumber \\
&&+2\bar{h}(x)^{ab}\bar{h}(x)_{ab}\Big( -2m^4 \Big)
\Big(Y+\frac{3}{2}\Big) \Bigg]    ,
\end{eqnarray}
\begin{eqnarray}
&L_{1}[&h_{ab}]=\frac{1}{512\pi^2}\int d^4 x \int d^4 x'\Bigg\{
\Big[\bar{h}(x)\bar{h}(x') \nonumber \\
&& \ \ \ +2\bar{h}^{ab}(x)\bar{h}_{ab}(x')\Big]\nonumber \\
&&\times\int \frac{d^4 k}{(2\pi)^4}e^{ik(x-x')}\Bigg[
\Big(Y+\frac{46}{15}\Big)
\Big(m^4+\frac{m^2}{3}k^2 +\frac{k^4}{30} \Big)   \nonumber \\
&&-\frac{m^4}{2}-\frac{m^2}{15}k^2 -\frac{1}{15}(k^2+4m^2)^2 F_1(k)
\Bigg]\Bigg\}   ,
\end{eqnarray}
\begin{eqnarray}
&L_{2}[&h_{ab}]=L_{3}[h_{ab}] \nonumber \\
&=&\frac{1/2}{512\pi^2}\int d^4 x \int d^4 x'\Bigg\{
\bar{h}(x)\bar{h}(x') \nonumber \\
&&\times \int \frac{d^4 k}{(2\pi)^4}e^{ik(x-x')}\Bigg[
\Big(Y+\frac{11}{3}\Big)
\Big(-4m^4-\frac{2m^2}{3}k^2 \Big)   \nonumber \\
&&+\frac{4m^4}{3}+\frac{4}{3}m^2(k^2+4m^2) F_1(k)
\Bigg]\Bigg\}
\end{eqnarray}
and
\begin{eqnarray}
L_{4}[&h_{ab}&]=\frac{1}{512\pi^2}\int d^4 x \int d^4 x'\Bigg\{
\bar{h}(x)\bar{h}(x') \nonumber \\
&&\times \int \frac{d^4 k}{(2\pi)^4}e^{ik(x-x')}\Big[
(Y+4)(2m^4)-4m^4 F_1(k)
\Big]\Bigg\}  .
\nonumber \\
\end{eqnarray}
Here we have defined the constant
\begin{eqnarray}
Y=\frac{2}{4-n}+\ln\frac{4\pi}{\mu^2} -\gamma -\ln\frac{m^2}{\mu^2} ,
\end{eqnarray}
and the function
\begin{eqnarray}
F_1(k)=\sqrt{\frac{k^2+4m^2}{k^2}}
     \mbox{arctanh} \Bigg(\sqrt{\frac{k^2}{k^2+4m^2}}\Bigg)   .
\end{eqnarray}

The term which contains the differences between the in-out and in-in
effective action is
\begin{eqnarray}
\label{U-formula}
U[&h_{ab+}&,h_{ab-}]=\frac{1}{256\pi^2}\int d^4 x \int d^4 x'
\int \frac{d^4 k}{(2\pi)^4}e^{ik(x-x')}  \nonumber \\
&&\times \Bigg\{
\bar{h}_{+}(x)\bar{h}_{-}(x')
\Big[-4m^4+\frac{4m^2}{3}(k^2+4m^2) \nonumber \\
&& \ \ \ \ \ \ \ \ -\frac{1}{15}(k^2+4m^2)^2 \Big]G_1(k)
\nonumber \\
&&+2\bar{h}_{+}^{ab}(x)\bar{h}_{ab-}(x')
\Big[-\frac{1}{15}(k^2+4m^2)^2 \Big]G_1(k)
\Bigg\}   ,
\nonumber \\
\end{eqnarray}
where
\begin{eqnarray}
G_1(k)=i\pi\sqrt{\frac{k^2+4m^2}{k^2}}\Theta(-k^2-4m^2)\Theta(-k^0) .
\end{eqnarray}


\begin{references}
\bibitem{explainlocal}
By ``local tensor'' is meant a tensor which is a functional of the
metric, and whose value at any point $p$ depends only on the metric in
an arbitrarily small neighborhood of $p$.
\bibitem{note1}
Further arguments based on dimensional analysis cannot eliminate the
ambiguous term (\ref{try2}): If one reverts to units in
which $\hbar \ne 1$ and $G \ne 1$, one can argue that the expected
stress tensor should be of the form
$$
T_{ab} = \hbar T_{ab}^{(1)} + O(\hbar^2),
$$
where $T_{ab}^{(1)}$ is independent of $\hbar$.  Also it can be argued
that $T_{ab}^{(1)}$ should be independent of $G$.  These requirements are
satisfied when one takes for $T^{(1)}$ the expression (\ref{try2}),
which has dimensions $({\rm length})^{-4}$.  [Note that in this
context $m$ is fundamentally an inverse lengthscale, not a mass scale,
from Eq.\ (\ref{class-matter})]
\bibitem{note4}
In addition, this postulate is incompatible with the result we find in
Sec.\ \ref{massless-limit} that the expected stress tensor predicted
by the in-in effective action formalism is not an analytic function of
$m^2$.
\bibitem{notepointsplit}
This ambiguity is presumably related
to a freedom in choosing the quantity conventionally denoted
$w_0(x,y)$ in the point splitting algorithm \cite{WaldQFT}.
\bibitem{note22}
Note that the axiom used in Ref.\ \cite{Horowitz} in the context of
linear perturbation theory --- that the stress tensor not depend on
derivatives of the metric of differential order higher than four ---
will not work outside of linear perturbation theory, as the ambiguous
tensor (\ref{try2}) satisfies this property.
\bibitem{note2}
Our example (\ref{try2}) of an infinite parameter ambiguity in fact is
not present in linear perturbation theory.
When one picks out the piece of the ambiguous contribution
(\ref{try2explicit}) to the
stress tensor that is linear in the metric perturbation, one finds
that the resulting tensor contains only the two parameter ambiguity.
However, there are other ambiguous terms with infinitely many
free parameters that are present at
linear order, for example terms generated by taking a variational
derivative of
$$
\int d^4 x \, \sqrt{-g(x)} \, m^2 F(\Box / m^2) R,
$$
where $F$ is a suitably chosen smooth function.
\end{references}
\end{document}